\documentclass[12pt,titlepage]{article}
\usepackage{times}
\usepackage{setspace}
\usepackage{amsmath}
\usepackage{amssymb}
\usepackage{graphicx}
\usepackage[round]{natbib}

\usepackage{epstopdf}

\DeclareGraphicsExtensions{.ps,.pdf,.eps}

\newcommand{\ee}{\end{equation}}

\newcommand{\magsim}{\ \lower2pt\hbox{$\sim $}\mkern-14mu \raise2pt\hbox{$>$}\ }

\begin{document}

\begin{titlepage}
\begin{center}
{\Large \bf A procedure to analyze nonlinear density waves in Saturn's rings using several occultation profiles}

\vskip 0.5truein
\baselineskip 16pt

{\large Nicole J. Rappaport$^\mathrm{*,1}$, Pierre-Yves Longaretti$^\mathrm{2,3}$,
Richard G. French$^\mathrm{4}$, Essam A. Marouf$^\mathrm{5}$, and Colleen A. McGhee$^\mathrm{4}$} \\ \vskip 0.5truein
{\small
$^\mathrm{1}$ Jet Propulsion Laboratory, California Institute of Technology, m.s. 301-150, 4800 Oak Grove Drive, Pasadena CA 91109 \\
$^\mathrm{2}$ Universit\'{e} Joseph Fourier, Laboratoire d'Astrophysique de Grenoble, BP 53, 38041 Grenoble Cedex 9 FRANCE \\
$^\mathrm{3}$ Centre National de la Recherche Scientifique (CNRS) / INSU \\
$^\mathrm{4}$ Astronomy Department, Wellesley College, Wellesley, MA 02181 \\
$^\mathrm{5}$ San Jose State University, One Washington Square, San Jose, CA 95192 \\
\bigskip $^*$ Corresponding author e-mail address: Nicole.J.Rappaport@jpl.nasa.gov} \end{center} \vfill
\noindent Revision: 16\\
\noindent Pages: 50 \\
\noindent Tables: 2 \\
\noindent Figures: 13 \\
\clearpage

\noindent \textbf{Proposed running head:} Density Wave Analysis

\vskip 2truecm

\noindent \textbf{Editorial correspondance to:}
\smallskip
\noindent Dr Nicole J.\ Rappaport \\
\noindent Jet Propulsion Laboratory \\
\noindent Mail Stop 301-150 \\
\noindent 4800 Oak Grove Drive \\
\noindent Pasadena, California 91109 \\
\noindent Phone: 818 354 8211 \\
\noindent Fax: 818 393 7116 \\
\noindent E-mail address: Nicole.J.Rappaport@jpl.nasa.gov \clearpage

\vglue 3truecm

\noindent \textbf{ABSTRACT} \\

Cassini radio science experiments have provided multiple
occultation optical depth profiles of Saturn's rings that can be
used in combination to analyze density waves. This paper
establishes an accurate procedure of inversion of the wave
profiles to reconstruct the wave kinematic parameters as a
function of semi-major axis, in the nonlinear regime. This
procedure is established using simulated data in the presence of
realistic noise perturbations, to control the reconstruction
error. It is then applied to the Mimas 5:3 density wave.

There are two important concepts at the basis of this procedure.
The first one is that it uses the nonlinear representation of
density waves, and the second one is that it relies on a
combination of optical depth profiles instead of just one profile.
A related method to analyze density waves was devised by
\cite{LB86} to study the nonlinear density wave associated with
the Mimas 5:3 resonance, but the single photopolarimetric profile
provided limited constraints. Other studies of density waves
analyzing Cassini data (\citealt{Colwell07, Tiscareno07}) are
based on the linear theory and find inconsistent results from
profile to profile. Multiple cuts of the rings are helpful in a
fundamental way to ensure the accuracy of the procedure by forcing
consistency among the various optical depth profiles.

By way of illustration we have applied our procedure to the Mimas
5:3 density wave. We were able to recover precisely the kinematic
parameters from the radio experiment occultation data in most of
the propagation region; a preliminary analysis of the
pressure-corrected dispersion allowed us to determine new but
still uncertain values for the opacity ($K\simeq 0.02$ cm$^2$/g)
and velocity dispersion of ($c_o\simeq 0.6$ cm/s) in the wave
region.

Our procedure constitutes the first step in our planned analysis
of the density waves of Saturn's rings. It is very accurate and
efficient in the far-wave region. However, improvements are
required within the first wavelength. The ways in which this
method can be used to establish diagnostics of ring physics are
outlined. \\

\bigskip
\noindent

\textbf{Key Words}: Saturn's rings, nonlinear density waves

\end{titlepage}

\clearpage

\setcounter{page}{5}

\section{Introduction}

The wealth of new data provided by the Cassini mission (and in
particular, the collection of independent radial wave optical
depth profiles of Saturn's rings) opens the possibility to use
wave dynamics to establish physical diagnostics of ring physics to
a level of accuracy substantially higher than what was possible
with Voyager. However, such a research program requires an
accurate determination of the wave kinematic parameters throughout
the wave propagation region. This has prompted us to reinvestigate
this last problem.

Many density waves in Saturn's rings are strong and nonlinear --
that is they cannot be modeled by the linear theory of density
waves fully described e.g. by \cite{S84} -- but this linear theory
is nevertheless used due to the absence of a well-established
nonlinear inversion procedure ({\sl e.g.}, \citealt{NCP90,
Rosen91a, Rosen91b, S04, Tiscareno06, Tiscareno07}). This paper
presents a new approach to analyze linear and nonlinear density
waves. The procedure is established through the use of a number of
simulated optical depth profiles, as only simulated data allow us
to compare the reconstructed kinematic parameters with the
original ones. It is applied next to the Mimas 5:3 density wave as
observed by the Cassini Radio Science occultation experiment, as
an example. This procedure will serve as the basis of several
subsequent studies of real data. The approach adopted here
considerably sharpens the method described by \cite{LB86}, which
extracted the maximum information from one Voyager
photopolarimeter stellar occultation profile. However, one profile
is insufficient to accurately constrain all the parameters of the
model to the level of accuracy required to produce new and
detailed diagnostics of the ring physics.

There are several motivations for analyzing density waves, and
especially nonlinear density waves. The first objective of
determining the kinematic behavior of ring density waves is to
test the background kinematic model described in section
\ref{theory}, as possible systematic deviations from this model
might provide information on physics that are yet unknown or not
well constrained. Cassini provides many different radio
occultation profiles of the rings, which can be used to verify
that a single set of wave parameters can be determined for various
observations of a wave at various longitudes with respect to that
of the associated satellite (at least when modulations due to
satellite orbital variations or other physical effects are weak
enough).

For each wave, the set of kinematic parameters determined by the
method described in this paper, coupled to the analysis of the
evanescent part of the wave (which may require extraneous
dynamical constraints to be accomplished), form the basis of the
next objective, which is to measure as precisely as possible the
torques exerted by the satellites on the rings. Such a direct
measurement was previously attempted in \cite{LB86}, but with
limited success due to insufficient constraints in the first
wavelength of the studied wave, and to failure of the WKBJ
approximation in the vicinity of the resonance. This type of
torque measurement based on a determination of the wave kinematics
can only be performed in Saturn's rings, and therefore constitutes
a unique way to verify the dynamical theoretical predictions for
such torques. The measurement and verification of the torque are
directly relevant to the dynamics of disk-satellite interactions
in general. The extent to which dynamical constraints are required
on top of a purely kinematic description of the wave remains to be
seen; this bears on the model dependence of such a torque
measurement.

Another important study that will follow from this paper is the
detailed analysis of the ring stress tensor, with the hope that
this may in turn provide useful constraints on the ring particles
collisional properties. We wish to identify statistically the
various stress behaviors that might be found, depending, for
instance on the ring region or ring background optical depth.
Several models of damping exist. For instance, \cite{BGT83} and
\cite{SDLYC85} predict a bimodal behavior as a function of the
ring mean optical depth in dilute rings. \cite{BGT85} generated
models for dense rings. This research program requires combining
the kinematic reconstruction method developed here with the
nonlinear (pressure-corrected) dispersion relation and at least
one generic dynamical equation describing wave propagation and
damping (such as the ones derived in \citealt{SDLYC85} or
{\citealt{BGT86}).

In the bulk of this paper, we simulate a density wave that has
some similarities with the Mimas 5:3 density wave, as observed in
eight diffraction-corrected Cassini radio occultation profiles. We
did not try to simulate this wave precisely, because the purpose
of the simulation was to establish the procedure. A preliminary
and simplified analysis of real data pertaining to this wave is
presented in section \ref{mimas}. The reader might wonder why we
started with simulated data. The reason is that such an analysis
is in itself already very complex and it is important to separate
this complexity from that inherent to the analysis of real data.
There are several reasons for which the analysis of simulated data
is difficult.

The first reason is the nonlinearity in itself. The equation
describing the shape of the wave, with its large troughs and
narrow peaks, does not provide a model that one can readily fit to
the data. The fit turns out to be extremely sensitive to the
dependence of the wave phase on semi-major axis, and so the
procedure must allow for a very precise determination of this
quantity. By using simulated data, we can compare how the
reconstructed wave kinematic parameters such as the wave phase
compare to the ones used in constructing the data; we can
therefore quantify the error in the reconstruction method
developed here.

The next difficulty is associated with the fact that the absolute
radial scale of Saturn's rings is presently known to only $\pm 2$
km (\citealt{NCP90, French93, Jacobson06}). Long period variations
in the satellite's orbit may produce similar radial shifts in the
resonance location, and are expected to produce variations in the
wave propagation which are not yet modeled. The uncertainty in the
radial scale, at least, will be reduced by future kinematic models
to the level of about 100 m.

Another major problem comes from the fact that one must
distinguish between radius $r$ and semi-major axis $a$; the
relation $r=a\left[1-e\cos(m\phi+m\Delta)\right]$ (see section
\ref{theory} for definitions of the variables and discussion of
this relation) is simple only superficially.

Other effects add their own complexity to the data. Interestingly,
the noise present in real data (which we took into account in our
simulations) is not a real problem, although the need for several
profiles to constrain a unique self-consistent solution results
mostly from the presence of noise. However, and as indicated by
the work of \cite{LS00, Lewis05}, gravitational clustering can
cause significant effects on the structure of local density
maxima. Also, the gravitational wakes observed in the rings (see
for instance \citealt{Hedman07, Colwell06, Colwell07a}) make it
necessary to normalize the background optical depth among the
various profiles, and may affect the optical depth profile of the
wave in an as yet uncontrolled way. Focusing on simulated data
allows us to ignore this added complexity in the first step .

Before describing our procedure in section \ref{procedure}, we
begin in section \ref{theory} by gathering the basic equations of
the theoretical streamline kinematic model. Section
\ref{simulation} describes and illustrates the method to obtain
the simulated data. Section \ref{mimas} applies the procedure to
the Mimas 5:3 density wave. Section \ref{conclusion} summarizes
and concludes the paper with a discussion of future work.

\section{Theoretical Model}\label{theory}

The most striking observed characteristic of density waves -- the
shape of the peaks and troughs -- is kinematic in nature. Dynamics
enters mostly through the nonlinear dispersion relation, which
specifies the change of the wavelength with radius as the wave
propagates, and through the control of the wave amplitude; another
dynamical feature is the existence of an evanescent zone. More
generally, dynamics controls the way the observed kinematic
features are spatially modulated. It is therefore legitimate first
to focus on the information that can be recovered from kinematic
constraints, relying as little as possible on dynamical ones.

The major difficulty of this program lies in the fact that the
observed wave profiles depend in a complex, mostly implicit way on
the underlying kinematics, making the elaboration of an inversion
procedure both a necessary and complex task. In this section, we
gather the required kinematics (and the minimal dynamics) needed
to fulfill this objective.

This section addresses test particle kinematics (section
\ref{kinematics}), streamlines (section \ref{streamlines}),
surface density and optical depth (section \ref{sigmaandtau}), and
finally discusses the link between radius and semi-major axis
(section \ref{randa}), which plays an important role in the
inversion procedure. Particular attention is paid to the
discussion of the generality of the kinematic model and dynamical
constraints used in our inversion procedure.

\subsection{Test Particle Kinematics}\label{kinematics}

We consider a satellite orbiting in the equatorial plane of the
planet. The cylindrical coordinates are denoted by $(r,\theta)$,
where $r$ is the radius and $\theta$ is the longitude. We use the
subscript $s$ to characterize the satellite.

As in \cite{Goldreich82}, one can linearize the equations of
motion of a test particle orbiting the planet to obtain

\begin{eqnarray}
\frac{d^2r}{dt^2} - r \left( \frac{d\theta}{dt} \right)^2 &=&
-\frac{\partial \Phi}{\partial r} \qquad {\rm and}
\label{eqmotionone} \\ \frac{d}{dt} \left( r^2 \frac{d\theta}{dt}
\right) &=& - \frac{\partial \Phi}{\partial r},
\label{eqmotiontwo} \end{eqnarray}

\noindent where

\begin{equation}
\Phi=\Phi_p+\Phi_s \label{potential}
\end{equation}

\noindent is the sum of the planet potential and the satellite
potential, to obtain the first order solution for $r$ and
$\theta$.

At any given time $t$, the satellite potential exerted on a
particle at $(r,\theta$) is periodic in $(\theta-\theta_s)$.
Furthermore, the satellite makes an epicyclic oscillation at
angular frequency $\kappa_s$, the epicyclic frequency, about a
guiding center which revolves at the rate $\Omega_s$, the mean
motion. Thus the satellite potential can be expanded in a double
Fourier series, one in longitude and one in time:

\begin{equation}
\Phi_s(r,\theta,t)=\sum_{m=0}^{+\infty}\sum_{k=-\infty}^{+\infty}
\Phi_{mk} \ \cos\left[ m(\theta-\theta_s) +k \kappa_s (t-t_0)
\right], \label{satpot}
\end{equation}

\noindent where the integers $m$ and $k$ characterize each term of
the series and $t_0$ is the time of satellite periapse. The
pattern speed

\begin{equation}
\Omega_p=\Omega_s+\frac{k}{m} \kappa_s, \label{patternspeed}
\end{equation}

\noindent is the angular speed of the rotating frame in which the
$(m,k)$ component of the satellite potential is stationary.

Denoting by $\Omega_0$ the angular rate of the ring particle in
the zero order solution of Eqs. (\ref{eqmotionone}) and
(\ref{eqmotiontwo}) for $r$ and $\theta$, the first order solution
is singular if $\Omega_p = \Omega_0$ (corotation resonance) or if
$m (\Omega_0 - \Omega_p)=\pm \kappa$ (Lindblad resonance). Inner
Lindblad resonances correspond to the $+$ sign and occur in a ring
inside the satellite, and outer Lindblad resonances correspond to
the $-$ sign and occur in a ring outside the satellite . We will
restrict ourselves here to inner Lindblad resonances. The
resonance is labeled $(m+k):(m-1)$ because we have

\begin{equation}
\frac{\Omega_0}{\Omega_s} \simeq \frac{m+k}{m-1},
\label{approximatecondition} \end{equation}

\noindent an approximation that we note here but will not use in
the remainder of this paper.

Near a Lindblad resonance, the linear response of a test particle
is given by:

\begin{eqnarray}
r &=& r_0+\frac{A_r }{\Delta r} \cos m \phi, \label{solutionone}
\\ \theta &=& \theta_0 + \Omega_0 t + \frac{A_{\theta}}{\Delta r}
\sin m \phi, \label{solutiontwo}
\end{eqnarray}

\noindent where $\Delta r$ is the distance from the resonance and

\begin{equation}
m \phi=m\left[ \theta-\theta_{s0} - \Omega_p (t-t_0) \right],
\label{mphi} \end{equation}

\noindent with

\begin{equation}
\theta_{s0}=\theta_s(t_0). \label{thetaszero}
\end{equation}

The coefficients $A_r$ and $A_{\theta}$ are given explicitly in
\cite{Goldreich82}.

As noted earlier, the amplitude of the first order solution for
$r$ and $\theta$ becomes infinite at the resonance location. The
linearization assumption that the first order solution is much
smaller than the zeroth order solution breaks down too close to
the resonance.

\subsection{Streamlines}\label{streamlines}

When considering \textit{fluid} particle motions instead of test
particle ones, it turns out that collective interactions
(collisions and self-gravity) prevent the divergence just
mentioned from occurring; they also make the description in terms
of first order deviations from circularity valid to a high level
of precision (relative deviations are typically of the order of
$10^{-5}$ or $10^{-4}$ in observed density waves, an order of
magnitude that can be deduced either from linear theory or from
order of magnitude analysis of the data; see \citealt{LB86} for
details), while changing the amplitude and phase of this first
order description. Because waves are forced by external satellites
and because self-gravity acts as a cohesive force on wave motions,
fluid particles sharing the same semi-major axis are expected to
follow the same $m$-lobe orbit, or streamline, in the frame
rotating with the pattern speed associated with the considered
satellite resonance [see Eq. (\ref{patternspeed})]. The shape of
ring streamlines is given by:

\begin{equation}
r = a \left[1 - e(a) \cos \left( m \phi + m \Delta(a) \right)
\right], \label{streamlinestwo}
\end{equation}

\noindent where $a$ is a semi-major axis, $e(a)<<1$ is the
eccentricity, and $\Delta(a)$ is a lag angle; as in the previous
section, $\phi$ is the azimuth in the rotating frame. The lag
angle has a direct geometric interpretation: it is the angle by
which the m-lobe shape of the wave rotates when moving from one
streamline to the next [see Fig.~\ref{fig0} where $\psi$ is
defined by Eq.~(\ref{psi})].

This description implicitly assumes a Lagrangian approach to fluid
motions. An unperturbed fluid particle follows a circular orbit,
and has coordinates $a,\phi$. Once perturbed, this same particle
follows an $m$-lobe orbit and has coordinates $r(a,\phi),
\theta(a,\phi)$, where $r(a,\phi)$ is the relation above and

\begin{equation}\label{thetaofphi}
\theta=\phi+2\frac{\Omega}{\kappa}e\sin(m\phi+m\Delta).
\end{equation}

In this description, $a,\phi$ are used as Lagrangian labels of
individual fluid particles, instead of the more customary initial
position $r_0, \theta_0$. It is important to note that $a,\phi$
are both the unperturbed position of the fluid particle, and its
Lagrangian labels. Functions of space, such as the ring surface
density, are either specified in terms of $r,\theta$ or, more
often, as a function of $a,\phi$ through the change of variables
defined by the two previous relations.

Eqs.~(\ref{streamlinestwo}) and (\ref{thetaofphi}) can either be
viewed as an \textit{a priori} assumption concerning fluid
particles kinematics, to be validated through the derivation of
appropriate dynamical solutions of the equations, an approach
adopted in \cite{BGT86}; or it can be derived from \textit{ab
initio} analysis of these same equations, as in \cite{SYL85}. Note
however that Eq.~(\ref{streamlinestwo}), the most important of the
two, constitutes in practice the most general such relation one
can assume: the fluid particle response must be periodic in
$m\phi$ because the satellite forcing is\footnote{Only the lowest
terms in eccentricity are important.}; and the magnitude of the
observed deviations from circularity ensures that a sinusoidal
periodic response (the first term expansion of any periodic
response) will accurately represent the data to their current
level of precision.

For our purposes, Equation~(\ref{streamlinestwo}) involves several
implicit features. The first one is that the streamlines do not
cross. Non-interacting streamlines would cross at locations where
$\partial r/\partial a=0$; indeed, wherever a difference in
semi-major axis $\delta a$ would not produce a difference in
radius $\delta r = \partial r/\partial a\delta a$ at a given
azimuth, fluid particles with different semi-major axes would
occupy the same location, leading to streamline crossing. One
has\footnote{The name $J$ stems from the fact that $\partial
r/\partial a$ is the leading order term in eccentricity in the
expression of the Jacobian of the change of variable from
$r,\theta$ to $a,\phi$.}:

\begin{equation}
J=\frac{\partial r}{\partial a}_{\mid{\phi}} = 1 - q \cos \left( m
\phi + m \Delta + \gamma \right), \label{Jacobian}
\end{equation}

\noindent where we have neglected the small term $e
\cos(m\phi+m\Delta)$ and with:

\begin{eqnarray}
q \cos \gamma &=& a \frac{de}{da}, \label{qcosgamma} \\ q
\sin\gamma &=& m a e \frac{d \Delta}{da}. \label{qsingamma}
\end{eqnarray}

In Eq.~(\ref{Jacobian}), the lag angle between streamlines is
defined by Eq.~(\ref{streamlinestwo}) and $\gamma$ and $q$ are
defined by Eqs.~(\ref{qcosgamma}) and (\ref{qsingamma}).
Equation~(\ref{Jacobian}) is a mathematical consequence of
Eq.~(\ref{streamlinestwo}) and requires no extra physical input.
Streamline crossing is prevented as long as $q < 1$. It is known
from observations that although $e$ is a very small parameter, $q$
is usually a significant fraction of unity. The physical meaning
of $q$ and $\gamma$ will be further illustrated in the section
devoted to the surface density, \ref{sigmaandtau}.

From a dynamical point of view, as all forces are small compared
to the planet attraction, fluid particles must follow perturbed
epicyclic orbits, very much like test particles follow perturbed
elliptic (or possibly epicyclic) orbits. This fluid particle
epicyclic theory (see \citealt{BL87}, \citealt{LB91}, and
\citealt{Borderies94}) was devised to put the streamline formalism
on a more self-consistent footing. For an \textit{ab initio}
construction of the streamline formalism, see \cite{L92}. We have:

\begin{equation}
r=a \left[ 1- e \cos(\theta-\varpi) \right], \label{epicyclic}
\end{equation}

\noindent with:

\begin{equation}
\varpi=\varpi_0+\dot{\varpi}(t-t_0), \label{varpi}
\end{equation}

\noindent where $\dot{\varpi}=\Omega-\kappa$ is the fluid particle
apse precession rate.

Equations (\ref{streamlinestwo}) and (\ref{epicyclic}) are
compatible if (to lowest order in eccentricity):

\begin{equation}
(m-1)\theta_0-\varpi_0-m\theta_{s0} - m \Delta = 0,
\label{conditionone}
\end{equation}

\noindent and:

\begin{equation}
m(\Omega - \Omega_p) = \Omega - \dot{\varpi}. \label{conditiontwo}
\end{equation}

Equation (\ref{conditionone}) expresses the constraint that all
fluid particles with the same semi-major axis belong to the same
$m$-lobe streamline. This is a generalization of the condition
that an elliptical ring is such that all the particles with the
same semi-major axis have the same periapsis angle.

Equation (\ref{conditiontwo}) looks like the resonance condition,
but it includes all the perturbations exerted on the particles
(whose dominant contributions is on the precession rate), in line
with the general philosophy of osculating elements; as such it
applies throughout the wave region (see \citealt{BGT86}). This
description is useful as long as all these perturbations (arising
from interparticle collisions and self-gravity, mostly) are much
smaller than the effects of the planet, a condition largely
satisfied in ring systems. As a consequence, the streamlines do
not differ much from the orbits of test particles viewed in the
rotating frame. Note that as long as explicit expressions of the
forces are not used, the dynamical content of
Eq.~(\ref{conditiontwo}) is absolutely generic, as the osculating
epicyclic theory is mathematically equivalent to the generic form
of the equations of fluid motion in the limit of small deviations
from circularity. As such, this condition is simply a kinematic
consequence of Eq.~(\ref{streamlinestwo}).
Equation~(\ref{thetaofphi}) is valid to the same degree of
generality.

The angle $\Delta$ [defined by Eq.~(\ref{streamlinestwo})] is
introduced because the streamline orientation varies smoothly from
the inside to the outside of the resonance. Indeed, in Eq.
(\ref{conditiontwo}), the secular drift of $m(\Omega - \Omega_p) -
\kappa$ that would be due to the planet alone can be compensated
by self-gravity (a condition that by itself yields back the wave
dispersion relation; see, e.g., \citealt{BGT86} and
\citealt{L92}). But since self-gravity is a small force, this
compensation takes place only if the shift of $\Delta$ between
streamlines is large enough, {\sl i.e.} if we are in the condition
of the tight-winding approximation:

\begin{equation}
m a \frac{d \Delta}{da} >> 1, \label{tight-winding}
\end{equation}

This is why density waves are so tightly-wound in rings. In spiral
galaxies, the self-gravity dominates the force due to the central
bulge, radial excursions are much larger (so that a first order
deviation from circular motions such as the one adopted here is
much less appropriate) and the spiral arms are much more open. The
equations describing linear waves can be recovered the limit where
$q << 1$ (\citealt{LB86}).

Another useful qualitative dynamical feature relates to the WKBJ
ordering, which is usually assumed in theoretical analyses of
density waves. Let us point out the meaning of this approximation
for our purposes. A wave possesses three radial scales\footnote{A
fourth radial scale is obtained from $ae$. This scale is smaller
than any of the other three, and is not required in this
discussion.}: the semi-major axis, the wavelength $l$ (note that
the radial wavenumber $k\simeq m (d\Delta/da)$, a near-equality
that holds to leading WKBJ order), and the scale of variation of
the wave amplitude and background\footnote{For instance, the
lengths over which the amplitude of the wave increases and
decreases -- the damping scale, or the length of the wavetrain.},
which we will denote $\xi$. In order of magnitude, for a typical
wave, $a\sim 10^5$ km, $l \sim 10$ km, and $\xi\gtrsim 10 l$ (as
waves propagate for a few to a few tens of wavelengths). The WKBJ
ordering holds when the variation of the wave phase function
($m\Delta +\gamma$) with radius is much faster than the variation
of the wave amplitude, {\sl i.e.}, $k\xi \gg 1$.
Eqs.~(\ref{qcosgamma}) and (\ref{qsingamma}) imply then that
$q\cos\gamma\sim ae/\xi \ll q\sin\gamma \simeq kae$ and that
$\gamma\simeq \pi/2$ when the WKBJ ordering holds. Also, as all
density waves excited at inner Lindblad resonances display
decreasing wavelength and decreasing amplitude $q$ with increasing
radius (at least in the far propagation region for the amplitude),
and as $q\simeq k a e$, the decreasing amplitude is related to a
decrease of the eccentricity on scale $\xi$.

The WKBJ ordering ($k\xi \gg 1$) is satisfied in Saturn's rings,
albeit to a much lesser degree of precision than the tight-winding
approximation ($ka \gg 1$). This motivates us to assume that most
of the radial variation of the wave is due to the dependence of
the lag angle $\Delta$ on semi-major axis. We nevertheless devise
a correction to account for the imperfection of the WKBJ ordering,
as a precise determination of the phase function is essential to
the robustness of our procedure.

Note that both the linear (e.g., \citealt{S84} and references
therein) and nonlinear (e.g., \citealt{SYL85} and \citealt{BGT86})
density wave dispersion relations imply that $k\rightarrow 0$ at
the resonance radius, so that the WKBJ ordering \textit{must} fail
there and $\gamma\simeq 0$ in the evanescent part of the wave and
close to the resonance. This fact limits our ability to
reconstruct the wave kinematics inside the first wavelength, a
point that will be further discussed later on.

For future reference, we stress that several different ``phases"
are used in this paper, and introduce some terminology to
distinguish them. These phases are the lag angle $\Delta$, the
phase function defined by:

\begin{equation}
f(a)=m\Delta(a)+\gamma(a), \label{phase}
\end{equation}

\noindent and the profile phase $\psi$:

\begin{equation}
\psi=m\phi+f. \label{psi}
\end{equation}

The phase function is a characteristic of the wave ({\sl i.e.} all
the profiles have the same phase function) and the profile phase
characterizes the location of peaks and troughs in 2D as well as
in each 1D wave profile. The profile phase is the one which may be
associated to the usual concept of wave phase; e.g., the density
maxima in the ($a,\phi$) plane correspond to $\psi=2\pi$ modulo
$2\pi$ in the WKBJ ordering, while minima correspond to $\psi=\pi$
modulo $2\pi$ (see Fig.~\ref{fig0}). The phase function is a
useful intermediate quantity which gathers the semi-major axis
dependence of the profile phase, and whose derivative is, by
definition, the radial wavenumber.

\paragraph{Summary:}

This subsection (as well as the previous one) has introduced and
discussed a number of elements of wave kinematics and dynamics. In
our reconstruction procedure, the only kinematic relation
introduced here that we use is Eq.~(\ref{streamlinestwo}) and its
mathematical consequence Eq.~(\ref{Jacobian}), along with
Eq.~(\ref{thetaofphi}). We have argued that they are quite
general; furthermore, Eq.~(\ref{thetaofphi}) is used only in a
limited way, to justify that the difference between $\theta$ and
$\phi$ can be ignored, a consequence of $e \ll q$. The only piece
of dynamics used is the WKBJ ordering, {\sl i.e.}, the fact that
the radial variation of the lag angle $m\Delta$ is much larger
than the variation of amplitudes and background $e$, $q$, and
$\Sigma_0$ (the ring unperturbed surface density). As a
consequence $\gamma\simeq \pi/2$ over most of the wave. As
discussed, this assumption is expected to fail inside the first
wavelength, and our simulated data incorporate this feature.

\subsection{Surface Mass Density and Optical Depth}\label{sigmaandtau}

As the length scales of the observed wave pattern are much larger
than the local ring thickness, it is legitimate to adopt a
two-dimensional description. Conservation of the mass of a ring
element in the Lagrangian change of variable $r,\theta$ to
$a,\phi$ implies that $dM\equiv \Sigma(r,\theta) dr
d\theta=\Sigma_0(a)|J| da d\phi=\Sigma_0(a) da d\phi$ so that the
perturbed surface mass density given by:

\begin{equation}
\Sigma=\frac{\Sigma_0}{1 - q \cos \left( m \phi + m \Delta +
\gamma \right)}, \label{surfacedensity}
\end{equation}

\noindent where we have directly identified the Jacobian of the
change of variable with its dominant term in eccentricity given by
Eq.~(\ref{Jacobian}), while noting that $J > 0$ for $q<1$;
$\Sigma_0$ is the unperturbed surface density, also called the
background surface density. This follows because the ring mass
$\delta M$ enclosed between these two streamlines is the same
between perturbed and unperturbed states ($\delta M= \Sigma \delta
r\delta\theta = \Sigma_0 \delta a\delta\phi$), and because, to
leading order in eccentricity, the Jacobian of the change of
variable from $a,\phi$ to $r,\theta$ reduces to $J$.

This relation provides us with another meaning for $q$: at peaks,
$\sigma=\Sigma_o/(1-q)$ while at troughs, $\sigma=\Sigma_o/(1+q)$.
Therefore $q$ measures both the density variation due to the wave
propagation, and its nonlinearity, and $\gamma$ measures the
relative contribution of the variations in eccentricity and lag
angle to this density contrast [see Eqs.~(\ref{qcosgamma}) and
(\ref{qsingamma})]. Only the lag angle contribution to the density
contrast has been used in Fig.~\ref{fig0} (consistent with the
WKBJ ordering); this figure also illustrates that the variation of
distance of the streamlines is the direct cause of the surface
density variations of the wave, consistently with
Eqs.~(\ref{surfacedensity}) and (\ref{Jacobian}).

Application of Eq.~(\ref{surfacedensity}) to the ring normal
optical depth $\tau$ is usually taken for granted, but requires
some discussion. For simplicity, assume that the ring particle
distribution with respect to size, spin, shape, etc. can be
represented by a discrete index $i$, and let $N_i(r,\theta)$ be
the column density of particles of type $i$, $m_i$ their mass, and
$S_i$ their cross-section. With these definitions, the ring
surface density $\Sigma(r,\theta)=\sum N_i m_i$ while its optical
depth $\tau(r,\theta)=\sum N_i S_i$. If the distribution is
constant in time, the formal similarity between these two
relations implies that the optical depth $\tau$ will also formally
obey Eq.~(\ref{surfacedensity}), as mass conservation then reduces
to number conservation and as the wave restoring force (the
self-gravity) does not induce particle segregation. In such a
context, the optical depth is therefore also given by

\begin{equation}
\tau =\frac{\tau_0}{1 - q \cos \left( m \phi + m \Delta + \gamma
\right)}, \label{opticaldepth}
\end{equation}

\noindent where $\tau_0(a)$ is the "unperturbed" optical
depth\footnote{i.e., the optical depth that would be found if the
deviations from circular motions were suppressed, but this may
differ from the optical depth outside the wave propagation region;
see the following discussion.}.

The remarks above show also that the same relation will hold to a
high degree of precision if the size distribution, in particular,
is evolving slowly in comparison to the dynamical time-scale of
the wave. This depends in turn on the efficiency of gravitational
accretion in the ring region. \cite{Weidenetal84} put forward and
studied the concept of \textit{Dynamical Ephemeral Bodies}, or
DEBs. This concept leads to a particle size distribution evolving
on the time-scale of a rotation period $\Omega^{-1}$ \citep{L89},
but requires that the efficiency of gravitational accretion is
large and only mildly variable throughout a sizeable fraction of
the ring to account for the remarkably ubiquitous size
distribution found by \cite{Zebkeretal85}. However, a recent
reinvestigation of the efficiency of gravitational accretion by
means of numerical simulations shows that this is most likely not
the case, largely independently of particle collisional properties
\citep{KS04}. Because the dispersion velocity of ring particles is
so small, even in perturbed regions, this implies in turn that, on
average, only minute fragments of particles can be collisionally
eroded from or accreted onto ring particles, leading to very long
time-scales for the evolution of the particle size distribution,
validating the use of Eq.~(\ref{opticaldepth}), at least with
respect to this issue. Note that the particle-size distribution
within the wave may be different from the particle size
distribution outside the wave, because the collisional regime is
different. In this respect, the spectral ``halos" found by
\cite{Nicholson08} may indicate a variation in particle sizes in
the vicinity of strong density waves compared to other ring
regions. In any case, the presence of the wave certainly changes
the particle velocity dispersion and therefore the collisional
regime, leading to evolution of the particle size distribution on
long enough time scales throughout the wave region. This may be
accounted for by allowing the opacity $K=\tau/\Sigma$ to depend on
semi-major axis (any possible dependence on the azimuthal
direction being quickly erased by the ring velocity shear).

A related issue concerns the presence of self-gravity wakes, as
such structures form and dissolve on time scales comparable to the
orbital period (see also \citealt{KS04} and references therein).
It is unclear at this point how the presence of density waves
(which locally enhance or reduce the density) affect the existence
and properties of wakes, and how such wakes might affect the
optical depth profiles in the wave region. Also, wakes, and more
generally regions where the particles are closely-packed (as may
be the case in sections of the B ring or in density wave maxima),
certainly affect the way the optical depth is reconstructed from
the raw data.

These last issues are difficult to quantify. We therefore ignore
them for the time being, but we keep in mind that they may affect
in an uncontrolled way the application to real data of the
reconstruction method developed here.

In the simulated data presented here, we have assumed for
simplicity that the opacity $K=\tau/\Sigma=\tau_0/\Sigma_0$ is
constant, so that specifying $\tau_o$ as a function of semi-major
axis also specifies the radial dependence of the surface density.
We also ran simulations with a variable opacity, with no impact on
the precision of the reconstruction method.

\subsection{Kinematics, and the link between radius and Semi-Major Axis}\label{randa}

We have introduced a number of functions of semi-major axis:
$e(a)$, $\Delta(a)$, $q(a)$, $\gamma(a)$, $\Sigma_0(a)$,
$\tau_0(a)$. These are the fundamental kinematic functions that
our procedure aims at extracting from the data; they completely
specify the wave kinematics. Note that the recovery of $\Sigma_o$
from actual data requires some dynamical input, e.g., the
nonlinear dispersion relation; at the level of precision we aim at
in the end, uncertainties in the modeling of the ring stress
tensor may affect the reconstruction of $\Sigma_0$, a question we
will only briefly touch upon here.

Furthermore, we have used $a$ and $\phi$ as a set of spatial
coordinates; {\sl e.g.}, the surface density of the previous
section is spatially prescribed in terms of $a$ and $\phi$.
However, the data are given as a function of the radius $r$ and
azimuth $\theta$. Thus there is a need to apply transformations
from $a,\phi$ to $r,\theta$ and vice-versa.

One can check that assuming $\theta=\phi$ in such a change of
variable leads to negligible error at the level of precision
achieved in the data, due to the magnitude of the eccentricity,
and the way it comes into play in the change of variables (see
Appendix \ref{radiustosma}). Furthermore, in the radio occultation
profiles, the data acquisition procedure ensures that one can
assume that $\theta$ (and consequently $\phi$) is constant in a
given profile\footnote{This is not in general true for stellar
occultations.}, with negligible error at the level of precision of
the data (see next section). Therefore, it is legitimate to assume
$\Sigma(r,\theta)=\Sigma(r,\phi)$ (or a similar relation for
$\tau$), but one needs to specify how $\Sigma(a,\phi)$ is deduced
from $\Sigma(r,\phi)$.

Eq.~(\ref{streamlinestwo}) allows one to compute the radius $r$ as
a function of the semi-major axis $a$ and azimuth $\phi$ as long
as the functions $e(a)$ and $\Delta(a)$ are known; from our
previous remarks, $m \phi$ can be considered as a constant
parameter depending only on the optical depth profile considered,
so that this relation can be viewed as an $r(a)$ relation, and can
be inverted to find $a(r)$. The details of this inversion
procedure are discussed in Appendix \ref{radiustosma}.

Note that even though $ae$ is a small quantity compared to the
wavelength, it induces a difference between $\tau(a)$ and
$\tau(r)$, which can produce a significant error in the
determination of $q$ and $\tau_0$, and cannot be neglected.

\section{Simulated Data}\label{simulation}

In order for our simulated data to be realistic, we have extracted
some of the required parameters (such as longitudes and times)
from the diffraction-reconstructed occultation data for the eight
Cassini Radio Science Team occultation experiments
\citep{Kliore04} that will be used later on as a real test case of
our method. These data were acquired on days 123 (May 3rd), 141
(May 21st), 177 (June 26th), and 214 (August 2nd), respectively,
of 2005. On each day there was both an ingress and an egress
experiment, for a total of eight profiles. For each profile, the
data were acquired at the Deep Space Stations (DSS) of the Deep
Space Network DSS-14, DSS-43, DSS-43, DSS-63, DSS-14, DSS-14,
DSS-63, and DSS-63, respectively. The X-band data (frequency of
8.4 GHz) are plotted in blue in Fig.~\ref{fig7} with a resolution
of 1 km, with one point every 250 meters\footnote{The sampling
rate of the diffraction-limited data is 250 m, and when
diffraction is removed the effective spatial resolution is 1 km
(\citealt{Marouf86}).}, for the Mimas 5:3 density wave. This wave,
propagating from an inner Lindblad resonance location of 132301 km
from Saturn's center, is one of the strongest and most nonlinear
waves in the rings. It was particularly studied by \cite{LB86}.
These data have a free space signal-to-noise
ratio\footnote{Defined as $10 \log_{10} ($ averaged carrier power
/ average noise power in 1 Hz bandwidth $)$.} of 54 dB-Hz. The
instrumental noise is generally very low, although it is larger at
the peaks. Peaks that reach an optical depth of 5 are not physical
because they exceed the RSS capabilities; in other words, we have
no data points for the tops of these peaks. In addition to the
instrumental noise, the data also contain unmodeled physical
effects, such as peak shoulders and split peaks.

We consider each occultation of a given feature to be
instantaneous with its time $t$ being the mean time of the
occultation. This assumption introduces negligible error, as can
be checked by comparing the total data acquisition time for this
wave ($\sim$ 15 seconds for the Mimas 5:3 density wave) with the
dynamical time-scale associated with the pattern speed (18.1 hours
for the Mimas 5:3 density wave). We also assume that the ring
intercept point of the spacecraft-to-Earth ray occurs at a
constant mean inertial longitude $\theta$. This assumption is
justified for nearly diametric occultations, such as the ones
considered, which typically cover a longitude variation of
$0.01^{\circ}$. Therefore, we have one value of $m \phi$ for each
occultation. These values are shown in Table~\ref{tab1}.

We used these values of $m \phi$ and constructed eight simulated
occultation profiles in the following way. First we defined the
parameters $ae(a)$ and $\tau_0(a)$ using the array of radii of the
actual profiles as semi-major axes and the empirical equations:

\begin{eqnarray}
ae(a) &=& 0.1+ 2 \left[ 1+\tanh \left( \frac{a-a_c}{\zeta} \right)
\right] \left[\frac{a-a_N}{a_1-a_N} \right], \label{functionae} \\
\tau_0(a) &=& 0.5 + 0.5 \left[ 1+\tanh \left( \frac{a-a_c}{\zeta}
\right) \label{functiontauzero}\right] \left[\frac{a-a_N}{a_1-a_N}
\right], \label{functiontau0}
\end{eqnarray}

\noindent where $a_1$ (about\footnote{This value is in fact the
initial radius of our simulated data instead of its semi-major
axis. A similar comment applies to $a_N$.} 132260 km) is the first
(smallest) semi-major axis, and $a_N$ (about 132470) is the last
(largest) semi-major axis; $a_c=132275$ km and $\zeta=50$ km. We
selected $a_1=a_{res}$ as the resonance location for the simulated
data. These choices are made so that the resulting profiles
loosely resemble the observed wave. The choice of $\tau_0(a)$,
which increases at the beginning of the wave, is consistent with
the calculations of \cite{Shu85} and \cite{BGT86}, who find that
the surface density is enhanced in the wave zone outward of the
resonance. This is a general feature of strong waves.

Next, we defined the opacity $K$, and having the background
optical depth, this allowed us to compute the background surface
density $\Sigma_0$ in each point. As mentioned above, the
simulated data used in this paper have a constant opacity, which
is $K=0.0125$ cm$^2$/g. We also made tests with an opacity varying
with the semi-major axis.

The parameters $m \Delta$ and $q$ are next constrained in each
point by enforcing the ideal (WKBJ, pressureless, dissipation
free) nonlinear dispersion relation throughout the wave region.
This is certainly not satisfied in actual waves, but provides a
convenient way to test our ability to reconstruct all kinematic
parameters, including the surface density, in the wave propagation
region. The parameters $m\Delta$ and $q$ are computed iteratively
in the following way. Before starting the iterative calculation,
we computed in each point $q \cos \gamma$ from
Eqs.~(\ref{qcosgamma}) and (\ref{functionae}), and we initialized
$m\Delta(a_1)=0$. Next we computed:

\begin{equation}
m\Delta(a_n) = m\Delta(a_{n-1})+\int_{a_{n-1}}^{a_n} k da,
\label{compmdelta}
\end{equation}

\noindent where the wavenumber $k$ comes from the dispersion
relation \citep{BGT86}:

\begin{equation}
k=\frac{3(m-1)\Omega^2}{2 \pi G \Sigma_0 C(q)}
\frac{(a-a_{res})}{a_{res}}; \label{wavenumber}
\end{equation}

\noindent in this equation, $\Omega$ is the mean motion at
$a_{res}$, $G$ is the gravitational constant, and:

\begin{equation}
C(q)=\frac{4}{\pi} \int_0^{+ \infty} du \frac{\sin^2u}{u^2}
H\left(\frac{q^2 \sin^2u}{u^2}\right), \label{cq}
\end{equation}

\noindent with:

\begin{equation}
H(q^2)=\frac{1-\sqrt{1-q^2}}{q^2 \sqrt{1-q^2}}. \label{hqtwo}
\end{equation}

Having $m\Delta$, it is possible to compute $q \sin \gamma$ from
Eq.~(\ref{qsingamma}) and then $q$. We iterate the calculations
until the maximum change in the values of $q$ at the same point
and between iterations is less than 0.01. At the first iteration,
$C(q)=1$ (i.e., the dispersion relation reduces to the linear
one).

The parameter $\gamma$ is computed from Eqs.~(\ref{qcosgamma}) and
(\ref{qsingamma}); $f$ is then derived from Eq.~(\ref{phase}). The
simulated wave parameters are shown in Fig.~\ref{fig1}. Note that
$\gamma$ increases from $0$ to $\pi/2$ for the first thirty km of
the wave, so that in this region, our simulated data are not
consistent with the dynamics of a real wave (as the dispersion
relation fails there), although self-consistency of the kinematics
has been enforced in the way the simulated data have been
constructed. The simulated data nevertheless resemble an actual
wave in the first wavelength. The failure of the WKBJ ordering
inside the first wavelength is consistent with the findings of
\cite{LB86} for the Mimas 5:3 density wave. In this portion of the
wave and farther, $q$ increases as the wave becomes more and more
nonlinear until its growth (implied by self-gravitational
stresses) is stabilized to a plateau and then damped by viscous
stresses. Our simulated data also loosely reproduce this behavior.

Profiles without noise were derived from the above simulated
kinematic parameters, the values of $m \phi$, and
Eqs.~(\ref{streamlinestwo}), (\ref{psi}), and
(\ref{opticaldepth}). These profiles are shown by the blue lines
in Fig.~\ref{fig3}. The red lines on top of the blue lines
represent the reconstructed profiles, which will be discussed in
section~\ref{fpeaks}.

Next, we applied eight shifts in radii representing our imperfect
knowledge of the absolute radial scale in actual wave profile
recording. These shifts are $0, -0.5, -1.0, 0.5, 1.0, -1.0, -1.5,
1.5$ km. In other words, the first synthetic profile without noise
was not shifted, the second one was shifted by -0.5 km, the third
one by -1.0 km, etc. The ability to recover the values of these
radial shifts will be one test of our analysis method.

Next, we generated noisy data from the shifted profiles without
noise. For this, we used the actual data from 131980 km to 132190
km, {\sl i.e.} inward the Mimas 5:3 density wave profiles, in a
region where the optical depth fluctuations are somewhat muted. We
could have used free space data, but these data contain very
little noise. Our assumption was that whatever physical effects
give rise to optical depth fluctuations in the region inward of
the density wave plausibly acts also in the density wave. We used
the actual optical depths of each profile in the region considered
and inferred the amplitudes of the signal. For each profile, we
detrended the amplitudes by removing a parabola. We computed the
standard deviation of the detrended amplitudes. We generated an
array of pseudo-random values drawn from a normal distribution
with a mean of zero and standard deviation that is twice the
standard deviation of the detrended amplitudes. We averaged over
four consecutive samples to produce an array of amplitude noise.
This technique was used because the resolution of the data is 1
km, but they are sampled at 250 m intervals. We added this noise
to the amplitude of the simulated data without noise, and hence
generated noisy simulated data. The noise in these data is not
instrumental noise, it represents a real effect resulting from
real fluctuations in the optical depth (\citealt{Showalter90}).
The noisy profiles are shown by the blue lines in Fig.~\ref{fig4}.
Again, the red lines on top of the blue lines represent the
reconstructed profiles, which will be discussed in
section~\ref{fpeaks}.

The simulated profiles were interpolated to a common radial scale
extending from 132260 km (the assumed resonance radius) to 132470
km, with a step-size of 250 m as in the actual data. As mentioned
above, our goal was to generate plausible-looking profiles for
which the wave parameters are known, rather than accurately
replicating the Mimas 5:3 profiles. These simulated profiles were
used to establish the procedure which is described in the next
section. We chose the example of the Mimas 5:3 density wave to
guide us for generating the simulated profiles because this wave
is clearly visible and isolated in the data, and has previously
been studied by \cite{LB86}. The analysis of \cite{LB86} used only
one optical depth profile (a photopolarimeter profile), which did
not allow us to accurately constrain the location of the peaks,
the background optical depth, and the nonlinearity parameter $q$.
In this paper, we will apply the procedure to eight actual Mimas
5:3 radio optical depth profiles.

\section{Procedure}\label{procedure}

\subsection{Outline of the inversion procedure}\label{inversion}

Observations (or simulated data) provide us with
$\tau(r,\theta)=\tau(r,\phi)$ at specific azimuths $\phi$ (in the
frame rotating with the pattern speed). These observations are
modeled using $\tau(a,\phi)$ given by Eq.~(\ref{opticaldepth}),
using the $a(r)$ relation discussed in section~\ref{randa} and
Appendix~\ref{radiustosma}, so that
$\tau(r,\phi)=\tau(a(r),\phi)$. They involve three ``secondary"
functions of semi-major axis: the unperturbed optical depth
$\tau_0$, the nonlinearity parameter $q$ and the phase function
$f$ (Eq.~\ref{phase}). Furthermore, the quantities $q$ and $f$
depend on the ``primary" kinematic quantities, {\sl i.e.} the
eccentricity $e$ and the lag angle $\Delta$ (one can choose
$\gamma$ instead), through Eqs.~(\ref{qcosgamma}),
(\ref{qsingamma}), and (\ref{phase}). Our objective is to recover
the five functions $\tau_0(a)$, $q(a)$, $m \Delta(a)$, $ae(a)$,
and $\gamma(a)$ from the data. The major \textit{a priori}
difficulty lies in the fact that the relation between $\tau(r)$
and $\tau(a)$ at any given azimuth involves the knowledge of $e$
and $\gamma$, which are not known beforehand when fitting
Eq.~(\ref{opticaldepth}) to the data. This makes the recovery of a
common set of kinematic parameters for all profiles a challenging
task.

One possible method of inversion that would circumvent this
problem would be to perform a least-squares fit for each radius
(and at all radii at the same time), to the five previous
parameters with the help of the $\tau(a,\phi)$ and $a(r)$
relations. As one has more profiles than unknowns, this procedure
would seem to be well-posed, and would have a definite advantage:
no dynamical assumption of any kind (such as the WKBJ ordering)
would be required, only generically valid kinematic constraints
would be used. We have tried this, and, unfortunately, this fails
for several reasons. First, the number of profiles is limited, and
some of them have similar values of $m \phi$, so that they barely
provide enough independent information. For a simulation, we could
have used more profiles and a good sampling of values of $m \phi$,
but we wanted to see what we could do with the Radio Science data
that have been pre-processed so far. Secondly, the inversion
procedure is always extremely sensitive to errors in the
determination of the phase function $f$, so that, unless provided
with a guess unrealistically close to the actual phase function,
an iterative global least-squares procedure was found to fail
systematically. Thirdly, the radial origin of reference of all the
profiles is not the same, so that one must also fit for the radial
shifts one needs to apply to normalize the radial scales. In the
future we will know the radial absolute scale to within 100
meters, but again we wanted to see what we could do with the
existing Radio Science profiles. Note that even without radial
scale uncertainty, the resonance locations vary with time due to
the long period variation of the orbits of satellites such as e.g.
Mimas and Pandora. We found that radial shifts considerably worsen
the question of the precise determination of the phase function in
a global least squares fit. Fourthly, the smallness of the wave
amplitude and the loss of the WKBJ ordering in the vicinity of the
wave resonance makes the least-squares procedure less sensitive to
variations of $m\Delta$ in this region of the wave. Finally, the
relations between $q$ and $f$ on the one hand, and $e$ and
$\gamma$ on the other, involve a radial derivative; it is
difficult to have these derivatives numerically well-behaved in a
global fitting procedure. These difficulties make the
identification of a common set of kinematic parameters for all
profiles nearly impossible.  Nevertheless, a least-squares global
fit may be used as a final step to improve the quality of the fit,
a point we have not yet checked in any detail due to practical
implementation difficulties.

Instead, we have reverted to the approach devised by \cite{LB86},
with a number of significant improvements. The inversion procedure
is constructed from the iteration of two main stages. In the first
one, the secondary quantities $\tau_0(a)$, $q(a)$ and $f(a)$ are
deduced from the data for given primary parameters $e(a)$ and
$\gamma(a)$ (which provide us with a given $a(r)$ relationship);
this step relies on the WKBJ ordering, which we have shown to be
valid at least in the nonlinear part of density waves, but we
incorporate corrections due to the imperfection of this ordering.
In a second stage, $e(a)$ and $\gamma(a)$ are redetermined from
the $q(a)$ and $f(a)$ just obtained, with the use of
Eqs.~(\ref{qcosgamma}), (\ref{qsingamma}), and (\ref{phase}); the
WKBJ ordering is also used for convenience in this stage, but this
is not strictly necessary. Determination of the radial shifts of
the profile with respect to a common radial scale are interweaved
in these steps in a manner that will be specified below. These two
stages are iterated to improve the determination of $\tau_0(a)$,
$q(a)$ and $f(a)$ on the one hand, and $e(a)$ and $\gamma(a)$ on
the other. Iterations proceed until convergence is achieved. We
now turn to a more explicit and detailed description of these two
stages.

First, an initialization of all quantities is required before any
iteration is applied. This initialization and its rationale are
described in the next subsection.

Next, we present the first stage (recovery of the secondary
parameters), which is broken up into several steps. The first one
(subsection \ref{fpeaks}) describes the determination of the
secondary parameters $\tau_0$, $q$, and $\psi$ as a function of
the semi-major axis in the vicinity of the peaks for each of the
various profiles. This is the most complex step of our inversion
procedure; it relies on the WKBJ ordering. However, since a
precise determination of the phase function turns out to be
critical in our inversion procedure, we apply in this step a
correction to the leading order WKBJ solution for $\psi$ by taking
into account the small but non-vanishing variation of the wave
amplitude compared to the variation of the profile phase; this
correction might not always or everywhere be useful, but is
systematically computed, for simplicity. In the second step
(subsection \ref{fphase}), we determine the global phase function
$f$ from the profile phase $\psi$ as well as a first approximation
of the radial shift for each profile. We also make use of these
profile radial shifts to ``unshift" the data as well as $\tau_0$,
$q$, and $f$ such that they are on a common radial scale. The last
step determines the mean secondary parameters $\tau_0$, $q$ and
$f$ common to all profiles (subsection \ref{fmeansol}).

The second stage is divided into two steps: recovery of the
primary kinematic parameters $e$ and $\gamma$ (and coincidentally,
$m\Delta$) from the mean secondary parameters $q$ and $f$
(subsection \ref{faegam}), and final fitting of the reconstructed
profiles $\tau(r,\phi)$ on the data, which allows us to determine
an additional correction to the radial shift of the profiles
(subsection \ref{freconstruct}). This correction is needed because
first we are dealing with an iterative process with several passes
through the first and second stages, and second, we are making
assumptions that are only approximations, {\sl e.g.}, that
$\tau_0$ and $q$ are constant locally around each peak. The wave
kinematic parameters $\tau_0$, $q$, $f$, $ae$, and $\gamma$ are
shifted by the same amount as the profiles.

The WKBJ ordering is primarily needed in the first step of the
first stage, and, to the same degree of approximation, in the
recovery of $e$ and $\gamma$. The $a(r)$ relation is needed in the
first step of the first stage and in the last step of the second
stage. Also, as will be shown shortly, this ordering implies that
using only information at the profile peaks\footnote{The same
argument holds at troughs. However, peak locations are much more
precisely defined than troughs, although the experimental noise is
larger there; only peaks will be used in the procedure outlined in
this paper. The neglect of the troughs is compensated by the fact
that we have many profiles, whose peaks occur at different radii.}
or in the vicinity of the peaks minimizes errors in the knowledge
of $e$ and (small) deviations of $\gamma$ from $\pi/2$. Because
all five kinematic parameters we wish to recover vary smoothly
enough with semi-major axis, their determination at the peaks of
the profiles is sufficient to constrain them throughout the wave
region. Information is therefore gathered in the vicinity of peaks
only in the first steps of our procedure. This limited information
is sufficient to precisely reconstruct the wave kinematics, but
naturally, the goodness of the fit is checked throughout the wave
region by the reconstruction of the profiles.

In practice, the WKBJ ordering, as already mentioned, is not
robust enough in the first wavelength region (and in the
evanescent region as well); this is true of actual data, as found
by \cite{LB86}, and of our simulated data as well. As will be
seen, even in this region, our reconstruction method does not
entirely fail, but does not provide good enough results for our
ulterior purpose of torque measurement. Because of this
limitation, practical physical diagnostics of wave and disk
physics may require us to apply extraneous constraints to this
inner wave region. This point will be further discussed in our
conclusion section.

Our inversion procedure makes use of three different iterative
processes: one for the computation of $a(r)$ (see
appendix~\ref{radiustosma}), one for the recovery of $e$ and
$\gamma$ (see appendix~\ref{primaryparam}), and the entire
procedure itself. We found that, in general, three iterations of
the whole inversion process were sufficient to achieve convergence
to a high level of precision, with and without noise superimposed
on our synthetic profiles. Iterations of the whole inversion
procedure are referred to as ``passes" in the remainder of the
paper.

The reader may ask why such an involved inversion procedure is
required. We found by trial and error that simpler choices either
failed entirely, or produced unreliable results. This inversion
procedure has been implemented in MATLAB.

\subsection{Initialization}\label{initialization}

Our initialization is designed such that the kinematic parameters
can be treated as the results of a fictitious previous pass, for
implementation convenience. It assumes that the eccentricity is
equal to zero everywhere and that the profiles are not shifted.
Thus the semi-major axes are equal to the radii. As explained
below, we focus first on the peak regions, where $r=a$ holds to a
high level of precision, so initializing $e=0$ is a reasonable
choice to initiate the procedure. For the other parameters, we
initially assume $\gamma=\pi/2$ (strict validity of the WKBJ
ordering), $f=0$ (an arbitrary choice, but a better one would be
required only if $e\ne 0$, as it is only used in the $a(r)$
relation), $\tau_0=1$ (from the behavior of $\tau$ at the
beginning of the wave), and $q=0.5$ (midway between its possible
minimum and maximum values). For each profile, and for each peak,
the peak location is approximated initially by the semi-major axis
(or radius) of highest optical depth in a window of data
containing the peak.

\subsection{Recovery of the secondary parameters: profile phases and peak characteristics}\label{fpeaks}

We now come to the crucial problem of determining precisely the
peak locations as well as the corresponding $\tau_0$, $q$, and
$\psi$ of each profile using as initial estimates the results of
the previous pass (or of the initialization in the first pass).
This is one of the points on which we significantly improve on
\cite{LB86}. At each pass through our procedure, we first improve
the determination of $\psi(a)$ (subsection \ref{profphase}), and
then the determination of the peaks' characteristics (subsection
\ref{rqtau}).

In this step, we treat the various profiles independently.

\subsubsection{Profile phase determination:}\label{profphase}

Consistently with the WKBJ ordering, we note from
Figs.~\ref{fig2}, \ref{fig3}, and \ref{fig4} that the large-scale
variations of the wave parameters occur over a scale larger than
the wavelength. The variations in optical depths (see
Eq.~\ref{opticaldepth}) are therefore dominated by the variation
of $\psi$ (see Eq.~\ref{psi}) with semi-major axis.

Let us first look at what happens if this approximation is exact,
{\sl i.e.}, if $\gamma=\pi/2$ exactly throughout the wave region
and the variations of $q$ and $\tau_0$ with semi-major axis are
negligible; these approximations cannot hold exactly at the same
time (we will come to this below), but constitute a useful
starting point. In this case, Eq.~(\ref{opticaldepth}) implies
$\psi=2n\pi$ at peaks ($n$ is an integer, and successive peaks
correspond to successive values of $n$). It follows that $\cos
\left( m \phi+m \Delta \right)= \cos \left( \psi-\gamma \right) =
0$ and thus $r=a$ at the peaks, so that $\psi(a)=\psi(r)$ at the
peaks. We treat the resonance radius as a first ``peak" with $f=0$
(approximately true at resonance) so that $\psi=m\phi$ there
(modulo $2\pi$).

It is interesting to note that in the WKBJ approximation, the
value of the profile phases at the peaks is independent of the
peak locations; this is not quite the case if we take into account
the inaccuracy of the WKBJ approximation. To account for this
inaccuracy outside the first wavelength\footnote{Inside the first
wavelength, as the WKBJ ordering fails anyway, an altogether
different reconstruction procedure is required. See the discussion
section.}, we compute a correction with the help of the peaks'
characteristics determined in the previous pass (or of the
initialization in the first pass, and the correction vanished
then) in the following way.

Note that because $q$ and $\tau_0$ are not exactly constant, the
assumption that $\tau(r) \equiv \tau\left(a(r) \right)$ reaches a
maximum at points such that $r=a$ is incorrect for several
reasons. First, even ignoring the non-constancy of $\tau_0$ and
$q$, $\tau$ would indeed be maximum for $\psi=2n\pi$, but this
would not correspond to locations where $r=a$ because
$\gamma=\pi/2$ would not be exactly true. Secondly, there may be a
model error in that we have taken $\tau$ to the lowest order in
eccentricity in Eq.~(\ref{opticaldepth}). Thirdly, $\tau_0$ and
$q$ vary. Even though they vary slowly, these variations imply
that the maxima of optical depth do not correspond to
$\psi=2n\pi$. Of all three effects, only the third is possibly
non-negligible at the level of precision of the data, as can be
checked from order of magnitude estimates. The second is
negligible due to the very small eccentricities ($<10^{-4}$,
typically). The first is automatically taken into account in our
procedure in all passes but the first, although it is negligible
except close to the resonance. The third effect is accounted for
by computing a first order correction $\delta\psi$ to the phase
function, $\psi = 2n\pi + \delta \psi$, at the maxima of
$\tau(r)$, due to the non-vanishing derivatives $\tau_0' \equiv
d\tau_0/da$ and $q' \equiv dq/da$. To linear order in
$\delta\psi$, the constraint $d\tau/dr=0$ at the peaks semi-major
axis $a_{pk}$ yields:

\begin{equation}
\delta \psi(a_{pk}) = \frac{\tau_0'(a_{pk}) \left[ 1 - q(a_{pk})
\right] + \tau_0(a_{pk}) q'(a_{pk})}{q(a_{pk}) \psi'(a_{pk})
\tau_0(a_{pk})}. \label{deltapsi}
\end{equation}

This correction is appropriate except close to the resonance where
$\gamma \sim 0$.

\subsubsection{Improved determinations of the peak locations, and of $q$ and $\tau_0$ at the peaks:}\label{rqtau}

The procedure outlined in the previous subsection defines
$\psi(a_{pk})$ at the peaks of each profile, and $\psi(a)$ is
obtained from $\psi(a_{pk})$ by interpolation; this procedure is
quite accurate as $\psi(a)$ varies smoothly enough. Next, we make
use of this information to improve the determination of the
positions of the peaks, and to determine the values of $q$ and
$\tau_0$ at the peaks. To achieve this, we rely again on the fact
that these two quantities should be approximately constant in the
vicinity of any peak (or of any point for that matter) and find
them by least-squares fitting of the optical depth profile in the
vicinity of each peak with a function inspired from
Eq.~(\ref{opticaldepth}).

We first define an approximate optical depth profile function that
is valid around any peak $a_{pk}$ by

\begin{equation}
\tau_{fit}(a)=\frac{\tau_{0_{pk}}}{\left[1- q_{pk} \cos \psi(a)
\right]}, \label{taufit}
\end{equation}

\noindent where $\tau_{0_{pk}}$ and $q_{pk}$ are the values of
$\tau_0$ and $q$ at the considered peak, and $\psi(a)$ is the
profile phase function just determined. The use of an accurate
approximation to the true $\psi(a)$ turns out to be crucial for
the convergence of our inversion method. We then use this
approximate representation of the exact peak profile while
allowing for a possible error $\delta_{pk}$ in the peak location
by defining:

\begin{equation}
\hat{\tau}_{fit}(r) = \tau_{fit}\left( a(r+\delta_{pk}) \right),
\label{tauth}
\end{equation}

\noindent which amounts to a radial translation of $\tau(r)$. This
expression involves the knowledge of the function $a(r)$, which is
obtained from the inversion of $r(a)$ through the process
described in Appendix \ref{radiustosma} (in the first pass, $e=0$
so that $r=a$). This inversion uses the determination of $e$ and
$m\Delta$ obtained in the previous pass.

This function is then fitted by least-squares to the actual peak
profile in a window around the considered peak. This fit yields
the three unknown parameters of the function: $q_{pk}$,
$\tau_{0_{pk}}$ and $\delta_{pk}$. By trial and error, we found
that the window was optimal when allowing $\psi$ to vary by $\pm
2\pi/3$ around its peak value; this results from a compromise
between the validity of the assumption of constant $q$ and
$\tau_0$ around the peak, and the precision of the method.

When $e\neq 0$, $\tau_{fit}(r)$ is narrower in the peak regions
than $\tau_{fit}(a)$ which is a characteristic of the actual peak
profiles. Due to this property, our procedure results in
substantially improved values of $q_{pk}$ and $\tau_{0_{pk}}$ in
the second and subsequent passes with respect to the first. Note
that $\delta_{pk}$ is a peak radial position correction pertaining
to each peak, and is not related to the radial shifts applied in
section \ref{simulation} to mimic present inconsistencies in the
ring radial scale between actual profile data.

Finally, we update the peaks' semi-major axes to account for the
$\delta_{pk}$ corrections with the help of the $a(r)$ function, so
that, for each peak, we now have better approximations of $a_{pk},
\psi(a_{pk}), q(a_{pk})=q_{pk}, \tau_0(a_{pk})=\tau_{0_{pk}}$. The
full functions $\psi(a)$, $q(a)$ and $\tau_0(a)$ are specified by
interpolation.

\subsection{Recovery of the phase function and first radial scale correction}\label{fphase}

At this point in our procedure, we have determined the profile
phase $\psi(a)$ for each profile. What we need is a single phase
function $f(a)$, as precisely determined as possible due to the
sensitivity of the procedure to errors in this quantity.

First we compute the phase function for each profile by
subtracting $m \phi$ from $\psi$ [see Eq.~(\ref{psi})]. Next we
shift the phases by an integer number of $2 \pi$. Indeed, the way
we have ascribed values to $\psi$ at each peak does not ensure
coherence between profiles since peaks in different profiles
belonging to the same wavecrest may not correspond to the same
value of $\psi$, which may differ by an integer number of $2\pi$.
Once this is done, the various $f$ functions still present some
scatter, which is due both to errors in our reconstruction
procedure and to differences in the absolute radial scales of the
various profiles. We determine a first approximation of these
absolute radial scale differences by translating each profile by
an integer multiple of $2 \pi$ to minimize the difference between
the phases of the current translated profile and the first
profile. In this way we compute a single fixed radial shift for
each profile. We call this action ``collapsing the phases" because
if there were no approximation or assumption in our whole
procedure, all the phase functions would coincide after this
collapse. This determination is of course relative, since the
various radial shifts are computed by arbitrarily taking the first
profile as a reference profile. To compute the radial shifts, we
limit ourselves to the range from $a=132320$ km to $a=132440$ km,
where the phase is well-behaved and little affected by boundary
effects. The phases agree to a few percent in the range from
$a=132320$ km to $a=132440$ km, which we consider a very good
result.

The radial shifts (or more precisely ``unshifts") in the
semi-major axis (or radius) just determined must be accompanied by
similar unshifts of $\tau_0$ and $q$, $f$, and of the data
themselves. This requires some interpolation over an interval in
semi-major axis between $132270$ km and $132440$ km.

\subsection{Recovery of the secondary parameters: mean solution construction}\label{fmeansol}

We are now in position to use the functions $\tau_0(a)$, $q(a)$,
and $f(a)$ for each of the eight profiles to produce a single
function for each quantity that should be valid for all eight
profiles. In other words, we want to determine the mean solution
for the functions $\tau_0$, $q$, and $f$. We treat each of these
three functions separately.

It turns out that a polynomial fit to a simple average of each
function on each point does not satisfactorily represent the true
function ({\sl i.e.}, the function used to generate the simulated
profiles). Instead, we found that the following least square
procedure yields good results. The variables of this fit are the
mean values of the function over a limited number of points or
nodes (we chose 10 for this wave) approximately regularly spaced
between 132270 km and 132440 km, a range of values determined
empirically to avoid edge effects. These variables are first
estimated as the average values of the eight functions at each
node, providing initial conditions to the least square fit. The
function we wish to fit is computed everywhere by interpolation
over these 10 values, and is constrained to best represent, in the
least square sense, the set of eight profile-dependent functions
between 132270 km and 132440 km. In order to avoid negative,
unrealistic values of $q$, this quantity is constrained to a
small, positive value at the resonance radius. This somewhat
artificially constrains the behavior of $q$ in the first
wavelength. The negative values of $q$ derived from the various
profiles arise in part from the failure of the WKBJ approximation.

Figure \ref{fig2} shows the mean solution in red for $\tau_0$,
$q$, and $(f-f_{injected})/(2 \pi)$, on the six upper panels. The
left panels correspond to the case of data without noise, while
the right panels correspond to noisy data. The green lines refer
to the injected simulation parameters (i.e. the wave parameters
used to generate the simulated data and shown in Fig.~\ref{fig1}).
The blue lines show the values determined independently for each
profile. All the plots in the six upper panels are cut at 132270
and 132440 km because the failing of the WKJB approximation at the
beginning of the wave and boundary effects near the end of the
data set make the solution unreliable outside these cut-off
values. As a matter of fact, the solution is incorrect within the
first 50 km of the wave, where the wave function $f$ is not
determined accurately. As expected, there is more scatter in the
case with noise than in the case without noise. Nevertheless, the
agreement between the mean solution and the injected values
between 132310 and 132440 km is remarkable. This illustrates the
efficacy of this approach to find the mean solution for $\tau_0$,
$q$, and $f$. The four lower panels will be discussed in the next
section.

The improvement that this method brings to the determination of
$\tau_0$ and $q$, as compared to the individual determinations,
can be quantified by looking at Fig.~\ref{fig5}. The left panels
are for the case without noise and the right panels for the noisy
data. The top panel of this figure pertains to $\tau_0$ and the
bottom panel pertains to $q$. The middle panels will be discussed
in section \ref{recopacity}. In each panel, the magenta lines
display the function used to generate the simulated profiles (or
injected function). The green and cyan lines show the minimum and
maximum values $F_{min}$ and $F_{max}$, respectively, of the
computed $F=\tau_0$ or $F=q$ at any semi-major axis for the eight
profiles. The red lines represent the mean solutions computed in
this section. The black areas represent one standard deviation
error bars computed as (the ``noise" is statistically independent
in the various profiles):

\begin{equation}
\sigma_F=\frac{\left( F_{max}-F_{min} \right)}{2 \sqrt{N_{PROF}}},
\label{erbar}
\end{equation}

\noindent where $N_{PROF}=8$ is the number of profiles. We see
that for both $\tau_0$ and $q$, the magenta lines, which represent
the true solutions, fall within the error bar areas, and most of
the time coincide with the mean solutions, shown in red. Therefore
Eq.~(\ref{erbar}) provides a reliable handle of the error on
$\tau_0$ and $q$. Also this figure demonstrates the need to have
as many profiles as possible (preferably well distributed in $m
\phi$) to compute accurately $\tau_0$ and $q$. A precise
determination of these parameters is especially important to
determine the ring surface density and to constrain the ring
stress tensor, which will be done in a future paper.

\subsection{Recovery of the primary parameters: determination of $ae$ and $\gamma$}\label{faegam}

To solve for $ae$ and $\gamma$, we solve iteratively
Eqs.~(\ref{qcosgamma}) and (\ref{qsingamma}), a procedure
described in appendix \ref{primaryparam}. The equations are solved
between semi-major axes of $132290$ km and $132440$ km, as this
iteration procedure involved in the resolution relies on the WKBJ
approximation, which is not robust enough in the inner wave
region. Where needed in other subsequent passes, $ae$ and $\gamma$
are linearly extrapolated outside the range specified above. This
extrapolation is certainly one major source of error in this
reconstruction procedure, which explains the relatively large
(albeit tamed) difference between the injected and reconstructed
phase functions within the first wavelength region. Improvement is
clearly required on this point. This is discussed further in our
last section.

Figure \ref{fig2} shows the mean solution in red for $ae$ and
$\gamma - \pi/2$, on the four lower panels. As mentioned above,
the left panels correspond to the case of data without noise,
while the right panels correspond to noisy data. The green lines
refer to the injected simulation parameters. We find a remarkable
agreement between the mean solution and the injected values over
most of the wave. There is a difference in the first 50 or 60 km
of the wave, where the phase function was not well determined due
to the failure of the WKBJ approximation.

\subsection{Profiles reconstruction, and second correction to the radial scale}\label{freconstruct}

We are now in position to construct optical depth profiles
$\tau(r,\phi)=\tau(a(r),\phi)$ over the entire radial range of the
eight profiles, to check our obtained kinematic solution. We use
Eqs.~(\ref{streamlinestwo}) and (\ref{opticaldepth}). In doing so,
we determine by least squares a second approximation to the radial
shifts required by the inconsistent absolute radial scale of the
various profiles and the various approximations made in the
procedure. As before, this second shift translates into a
correlative shift of all our kinematic functions. The final radial
shifts are shown in Tables~\ref{tab1}.  The differences between
the applied and the computed data shifts are in general
significantly smaller in the absence of noise but even with noisy
data, they amount to at most several tens of meters, which is much
smaller than 250 m, the separation between the data points.

Figures~\ref{fig3} and \ref{fig4} show the unshifted reconstructed
profiles (in red) together with the unshifted simulated data (in
blue) for the profiles without (left panels) and with noise (right
panels). We can see that the agreement is excellent, and suggests
that a better determination of the radial scale shifts is hardly
possible. Note that the fit is still tamed inside the first
wavelength region, although the reconstruction of $ae$, $\gamma$,
and $m\Delta$ is only correct within a factor of two in this
region. Nevertheless, an improved determination of the kinematic
parameters will be required in this region to compute satellite
torques with some accuracy. The very last peaks are not accurately
represented because the fit was not performed in this region; the
reconstructed parameters have simply been extrapolated there.

\subsection{Recovery of the surface density}\label{recopacity}

Recovering the surface density requires some further dynamical
input, which is usually provided by the dispersion relation. As
the effect of the ring stress tensor is small, the pressure
correction is usually neglected in front of the self-gravity term
in the dispersion relation, which then reduces to the following
form in the region of validity of the WKBJ ordering:

\begin{equation}
K=\frac{2 \pi G C(q) k \tau_0 a_{res}}{3 (m-1) \Omega^2
(a-a_{res})}, \label{opacity} \end{equation}

\noindent where $K=\tau_0/\Sigma_0$ is the opacity. This
expression was applied in the construction of our simulated data.
The situation with real waves is more complex. \cite{Stewart08b}
have argued that the pressure term leads to measurable changes in
the dispersion on the order of 5\%.  For our application to the
Mimas 5:3 density wave, we have taken this term into account [see
Eq. (\ref{gth})].

The opacity is the quantity we chose to recover, as the optical
depth $\tau_o$, the nonlinearity parameter, and the wavenumber $k$
are known from the inversion procedure; $k$ is computed as the
derivative with respect to the semi-major axis of the phase
function $f$.

We also make use of the error estimates on $\tau_o$ and $q$ to
estimate the error on the recovery of the opacity $K$ and the
surface density $\Sigma_o$. In doing so, we can neglect the error
in $k$ because the error in the slope of $f$ is smaller than the
error in $f$, leading to relative errors on $k$ of the order of
one or two percent at most, much smaller than the error in the two
other quantities (of the order of 10 to 20\%).

We have found that the deviations in $q$ and $\tau_o$ of any given
profile with respect to the mean are not independent, but rather
tend to be anti-correlated. Therefore, they were not added.
Instead, we computed $K$ for five different cases, involving
several combinations of $\tau_0$ and $q$. Remember (see section
\ref{fmeansol}) that we denote by $\tau_{0,min}$ (or $q_{min}$)
and $\tau_{0,max}$ (or $q_{max}$) the minimum and maximum values
of $\tau_0$ (or $q$) in each data point computed independently for
each profile. We denote by $\overline{\tau}_0$ and $\overline{q}$
the mean values of $\tau_0$ and $q$ computed in section
\ref{fmeansol}.

The values of the opacity are shown in the medium panels of Fig.
\ref{fig5}. The left side corresponds to the data without noise,
and the right side to the noisy data. The nominal value of the
opacity, $K(\overline{\tau}_0, \overline{q})$ is displayed in red
while the injected value of $K=0.0125$ is plotted in magenta. In
the second panels from the top, the green lines show
$K(\tau_{0,min}, \overline{q})$ and the cyan lines show
$K(\tau_{0,max}, \overline{q})$. In the third panels from the top,
the green lines show $K(\overline{\tau}_0,q_{min})$ and the cyan
lines show $K(\overline{\tau}_0,q_{max})$. The error bars, drawn
in black, are computed by an equation similar to
Eq.~(\ref{erbar}). One can see that the true solution (magenta),
stays in the error bar area of the nominal solution between 60 and
170 km of the beginning of the wave. This demonstrates that the
procedure leads to a much better solution than that could be
inferred by considering the green and cyan lines, and that the
black area (one standard deviation) gives an accurate estimate of
the reconstruction error of the opacity.

The surface density can now be recovered easily as
$\Sigma_0=\tau_0/K$. It is plotted with red lines on Fig.
\ref{fig6}. The top panel is for the case without noise and the
bottom panel is for the case of the noisy data. The green lines
represents the injected value.

We compared our background surface density with what would be
predicted by the linear theory for each profile independently. For
each profile, we determined $k$ as the derivative of the profile
phase $\psi$ [see Eq.~(\ref{psi})], and we computed a constant
background surface density by a least-squares fit of the equation:

\begin{equation}
\frac{a-a_{res}}{a_{res} \Sigma_0}=\frac{2 \pi G k}{3 (m-1)
\Omega^2}, \label{lineardisp}
\end{equation}

\noindent which is the linear dispersion relation. This is similar
to what is routinely done in the literature.

The resulting values of $\Sigma_0$ for each profile are displayed
as blue lines on the two panels. These values are above the
average of the values obtained by using the nonlinear theory with
our procedure. Fig. \ref{fig6} clearly demonstrates the crudeness
of the linear theory when the surface density is not constant,
which is expected in real waves (see section \ref{simulation}).
Our ultimate goal requires that we recover the variation of the
surface density in the wave region; this makes the use of a
procedure of reconstruction of the wave kinematics such as the one
presented here a necessary first step.

\section{Application to the Mimas 5:3 density wave}\label{mimas}

The application of our procedure to the actual data collected for
the Mimas 5:3 density wave raises a number of other difficulties.
The first one is related to the variations in the orbital motion
of Mimas. Table \ref{tab2} lists the four resonance locations that
can be associated with the position of Mimas during the four
occultations which provide the eight profiles we have analyzed.
For each date, we have fitted over a period of two days the
epicyclic elements to the orbital data given by the SPICE kernel
{\sl sat267}, and then we have solved for the position of the
resonance. We see that the resonance location varies by 2.6 km,
i.e. a sizeable fraction of the (varying) wavelength, leading to
potentially non-negligible modulations of the wave between various
profiles. Two limiting cases can be considered: in the first, the
wave reacts in a quasi-static way to the variation of the orbit of
Mimas; in the other, the satellite orbital variations are fast
compared to the time-scale of adjustment of the wave. Orders of
magnitude estimates of the density wave group velocity show that
it is larger than the speed of variation of the resonance
location, but only by a factor of a few: the group velocity $\pi G
\Sigma / \kappa$ (\citealt{S84}) is approximately 20 or 30 km/year
for the Mimas 5:3 density wave, while the semi-major axis of Mimas
varies by 2 or 3 km with a period of 225 days (in addition to the
$\sim 70$ year period due to the resonance with Tethys).
Nevertheless, we deal with this difficulty by setting the
resonance radius at 132301 km and assuming one can absorb any
modulation of the wave in the radial shifts of the profiles, as
our aim is more to illustrate the behavior of our reconstruction
method than to derive precise quantitative results. We found
empirically that the radially rescaled profiles obtained in this
way are reasonably consistent with one another, indicating that
any spatial and temporal modulation of the wave behavior remains
mild enough, a conclusion consistent with the recent findings of
\cite{Stewart08a}.

The second difficulty, already mentioned in the introduction, is
that self-gravityl wakes cause the apparent background optical
depth to vary with viewing geometry
(\citealt{Colwell06,Hedman07}). We normalized the optical depths
of the eight profiles as follows. First we determined the mean
optical depth $\overline{\tau}_j$ for each profile in an
apparently unperturbed region of 20 km inward of the Mimas 5:3
density wave (see Table \ref{tab2}); it appears that modulations
in the mean optical depth are rather moderate. Thus, we computed
the average $\tau_{ave}$ of the $\overline{\tau}_j$ and we
multiplied the optical depth of each profile by
$\tau_{ave}/\overline{\tau}_j$. Nevertheless, the justification of
this procedure is unclear as we do not know if wakes inside the
wave, if present, affect the optical depth in the same way as
outside it. In any case, this is not a large effect, probably less
important than the uncertainty related to the large intrinsic
noise observed in the radio data.

Next, we applied our procedure to the optical depth data.
Fig.~\ref{fig7} shows the data (in blue) and the reconstructed
profiles (in red) after three passes. The agreement between the
model and the data is very good, considering the large
fluctuations of the optical depth (notably at the peaks) present
in the data. Fig.~\ref{fig8} is the equivalent of Fig.~\ref{fig2}
with a few differences. In the absence of injected value, the
phase function is compared to the its mean solution. The dotted
lines correspond to the region where the functions for at least
one of the profiles were extrapolated. In this extrapolation
region, the functions $\tau_0$ and $q$ increase. These functions
are nearly constant between 50 and 120 km from the beginning of
the wave (at 132301 km, the selected resonance radius).  In this
domain (132351 to 132421 km), the phase function $f$ is very well
determined. As expected, $ae$ increases and then decreases to
zero, and $\gamma$ is very close to $\pi/2$. Table \ref{tab2}
lists the shifts found for the eight profiles (they are all
positive, but this is only a coincidence).

Fig.~\ref{fig9} is the equivalent of Fig.~\ref{fig5} but does not
show the minimum and maximum values that were used to compute the
error bars, since they do not provide additional information. The
plots were traced in the domain where we can most trust the
solution, i.e., between 132350 km and 132420 km. As we have not
tried to determine a position-dependent opacity from the data, the
opacity is not shown in this plot.

Instead, the calculation of the opacity along with the velocity
dispersion is illustrated by Fig.~\ref{fig10}. This figure was
constructed as follows. We wrote the nonlinear dispersion relation
in the form:

\begin{equation}
g_{th}=g_{obs}, \label{nonlineardisp}
\end{equation}

\noindent with:

\begin{eqnarray}
g_{obs}&=&\frac{2 \pi G \alpha C(q) k \tau_0}{3 (m-1) \Omega^2},
\label{gobs} \\ g_{th}&=& \alpha K \left[
\frac{(a-a_{res})}{a_{res}} - \frac{2 k^2 H(q^2)
c_0^2}{3(m-1)\Omega^2} \right], \label{gth}
\end{eqnarray}

\noindent where $\alpha=1 \times10^{12}$ is an ad hoc numerical
factor designed to bring $g_{obs}$ and $g_{th}$ to values of order
unity and $a_{res}$ is the resonance semi-major axis. The second
term in Eq.~(\ref{gth}) is the pressure term, in the hydrodynamic
approximation where the (vertically integrated) pressure is given
by a simple isothermal equation of state $p=\Sigma c_0^2$; it
introduces the isothermal sound speed (of the order of the
particles' velocity dispersion) $c_0$ as a second parameter
besides the opacity in the dispersion relation. Indeed, we found
that the data were good enough to allow us to account for the
pressure correction to the dispersion relation, which leads to a better agreement
between $g_{th}$ and $g_{obs}$. At this
point, it is unclear if the remaining deviations from the purely
self-gravitational term are due to imperfection in the radio data
that would intrinsically bias the determination of $q$ and
$\tau_0$ (a point further discussed below), imperfections in the
model of the pressure term, neglect of the satellite term, or
position dependence of the opacity and velocity dispersion; note
in any case that \cite{LB86} pointed out that a constant opacity
provided a better fit to the dispersion relation than a constant
surface density.

We chose to determine only the opacity and velocity dispersion,
and not the resonance radius, for reasons also discussed below. We
find the velocity dispersion $c_0$ and the opacity $K$ by a
least-squares fit of Eq.~(\ref{nonlineardisp}) over the interval
between 50 and 120 km from the beginning of the wave, in the
following way.

In Fig.~\ref{fig10}, the solid lines represent the values of:

\begin{equation}
g_1=\alpha K \frac{(a-a_{res})}{a_{res}}, \label{defg1}
\end{equation}

\noindent and the dashed lines represent the values of:

\begin{equation}
g_2=\frac{2 \alpha \pi G \tau_0 k C(q)}{3 (m-1) \Omega^2} -
\frac{2 \alpha k^2 H(q^2) c_0^2 K}{3 (m-1) \Omega^2}.
\label{defg2}
\end{equation}

The black lines were obtained by using the mean values of $\tau_0$
and $q$. The red and blue lines on the top panel were obtained by
using the mean value of $\tau_0$ and the minimum and maximum
values of $q$ provided by the error bars on this variable. On the
other hand the red and blue lines on the bottom panel were
obtained by using the mean value of $q$ and the minimum and
maximum values of $\tau_0$ provided by the error bars on this
variable (one standard deviation). These error bars indicate that
deviations from strict linearity may be real; this point will be
further investigated elsewhere. In the nominal case, the solution
gives $c_0=0.55$ cm/s and $K=0.018$ cm$^2$/g.

Because of the non-constant background optical depth, the constant opacity results in a
non-constant surface density, shown in red in Fig.~\ref{fig11}.
The blue lines were obtained for the use of the dispersion
relation in the linear limit ($q \ll 1$). Note that we have $C(q)
\Sigma_{0,NL} ~ \Sigma_{0,L}$ where the subscripts $NL$ and $L$
refer to nonlinear and linear, respectively. Since $C(q)>1$, we
expect the nonlinear background surface density to be smaller than
the linear background surface density, as observed.

It is important to note that the cutoff of the radio data at an
optical depth of about 5 for the profiles considered, due to the
instrumental noise in the radio data, flattens the peaks to this
value. This affects the determination of $\tau_0$ and $q$. Also,
the unexpectedly large fluctations in optical depth observed at
the wave peaks with respect to troughs lead to a similar
systematic effect. Similarly, the noise level precludes a reliable
determination of the weak peaks in the wave damping region, which
is eventually needed to constrain the ring stress tensor. As
mentioned earlier, we have verified that the errors in $\tau_0$
and $q$ are anti-correlated. Attempts to fit
Eq.~(\ref{nonlineardisp}) with other combinations of $\tau_0$ and
$q$ led to the result that the best estimate of $a_{res}$ was
obtained for the minimum value of $\tau_0$ and the maximum value
of $q$. This indicated that $\tau_0$ is overestimated and $q$ is
underestimated, and led us to enforce the position of the
resonance in the fit. We conclude that the variations of the
background optical depth and surface density and the variations of
the nonlinearity parameter derived from the radio data are
imperfectly determined from the radio data. A better
determination, which could come from using optical occultation
data, is needed to study the damping of the wave and extract more
(and more reliable) information from the dispersion relation.

\section{Conclusion}\label{conclusion}

We have established a robust procedure to analyze nonlinear
density waves and we have applied it to the Mimas 5:3 density
wave. In principle, the number of independent profiles provided by
the Cassini mission would seem to make it possible to devise an
inversion procedure relying only on the kinematic characteristics
of the wave, but we found that in practice, and relying only on
the radio data, this fails for three major reasons: the
differences in the absolute radial scales of different profiles,
the complex way in which the $a(r)$ relation comes into play, and
the sensitivity to errors in the dependence of the wave phase on
semi-major axis.

Instead, we have extended and considerably improved the method
introduced in \cite{LB86}, which relies not only on the wave
kinematics, but on the WKBJ ordering as well. The intrinsic
nonlinearity of the problem makes the procedure complex.
Nevertheless, it is possible to reconstruct the wave parameters as
well as the data, in the presence of uncertainties in the absolute
scale of about $\pm 2$ km and of noise in the data; the data
resolution adopted was 250 m but the results should not be
sensitive to this parameter.

As such, this inversion procedure performs very well in the far
wave region. This opens new possibilities of physical diagnostics
of wave regions. In particular, with the help of the nonlinear
dispersion relation and the wave damping equation, one may
possibly constrain the surface density and ring's stress tensor in
a joint way, which would provide us with an indirect window on the
ring' collisional physics. We plan to make use of this inversion
procedure to undertake such systematic studies in the future.

The major inaccuracy of this reconstruction procedure is its
limited efficacy within the first wavelength of propagation of the
wave, where the WKBJ ordering is expected to fail on theoretical
grounds, an expectation confirmed by the behavior observed in
actual wave profiles. This limitation is most drastic in the
reconstruction of the eccentricity $e(a)$ and lag angle
$\Delta(a)$. This is unfortunate for the purpose of the
determination of the exchange of angular momentum with the
satellite, as the generic expression of the torque results from a
radial integral involving these two quantities [see, {\sl e.g.},
Eq.~(35) of \citealt{LB86}]. The results of the present inversion
method (which yields a fairly educated guess of the form of the
kinematic parameters even inside the first wavelength, and has
allowed us to remove the problem of inconsistencies of the
absolute radial scale between the various profiles), may ensure
the convergence of a refined, direct inversion method of the type
we initially tried, provided enough high quality profiles are
available. This approach may require to a different method of
inversion of Eqs.~(\ref{qcosgamma}) and (\ref{qsingamma}), if
needed; a possible way to do this is to re-express these equations
as a set of nonlinear algebraic equations, and solve them by an
algebraic iteration method. Additionally, further dynamical
constraints may be needed to ensure that such a direct inversion
method is well-defined. Three levels of such constraints are
available in the literature. In their most useful form, they are
all related to the dynamical equation governing the behavior of
$Z\equiv e\exp(i m\Delta)$ (see \citealt{SYL85}): they are the
dynamical equation itself, its solution in the linear limit, and
the asymptotic behavior of this solution in the evanescent region.
Defining the most appropriate strategy on this issue requires
substantial additional work, which will be reported elsewhere.

The application to the actual data of the Mimas 5:3 density wave
gives good results, although the determination of $\tau_0$ and $q$
is impaired by the noise in the radio data; better results are
expected from the use of, e.g., the UVIS data.

\appendix
\section{Appendix}
\subsection{Radius versus semi-major axis inversion procedure:}\label{radiustosma}

To compute the semi-major axis a a function of the radius we make
use of Eq.~(\ref{streamlinestwo}) in an iterative procedure. The
$(n+1)$th approximation of $a$, denoted $a^{(n+1)}$, is related to
the $n$th one through

\begin{equation}\label{rtoa}
a^{(n+1)}=r+a^{(n)}e(a^{(n)})\cos[m\phi+m\Delta(a^{(n)})].
\end{equation}

Convergence is ensured because $e \ll 1$; the procedure is
initialized with $a^{(0)}=r$.

The reader may ask why Eq.~(\ref{streamlinestwo}) cannot be
analytically inverted, at least approximately. This would be a
definite advantage in the numerical implementation of our
procedure, as the previous inversion is quite heavily required,
and slows down the whole process. In principle, one could do this
through an expansion in eccentricity. The zeroth order solution
is:

\begin{equation}
a = r + re(r) \cos (m \phi + m \Delta(r)). \label{zeroorder}
\end{equation}

We look for the first order solution by writing $a=r+\delta_1$
with $\delta_1={\cal O}(e)$. We find:

\begin{equation}
\delta_1=\frac{re(r) \cos (m \phi + m \Delta(r)) }{1-q(r) \cos \psi(r)}. \label{firstorder}
\end{equation}

The second order solution is of the form $a=r+\delta_1+\delta_2$
with:

\begin{equation}
\delta_2=\frac{1}{2} \frac{\left[ re(r) \cos (m \phi + m
\Delta(r))\right]^2}{\left[1 - q(r) \cos \psi(r)\right]^3} \left[
\frac{dq(r)}{dr} \cos \psi(r) - q \frac{d \psi(r)}{dr} \sin
\psi(r) \right]. \label{secondorder}
\end{equation}

Figure~\ref{fig12} shows the error between the approximate $a(r)$
obtained when using the zeroth, first, and second order solutions
and the actual $a(r)$. Naturally, the actual functions $e(a)$ and
$\Delta(a)$ (evaluated at $a=r$) are used in the process. Although
the eccentricity is very small, the error is substantial, of order
1 km, and the convergence is very slow due to the various powers
of the $1/J$ factors involved in this expansion, and more
crucially due to the fact that the order of the $k$th order term
in the expansion is not of magnitude $e^k$ but $ae(ae/\xi)^{k-1}$,
where $\xi$ is the scale of variation of the wave amplitude
introduced in section \ref{streamlines}; the factor $ae/\xi$ is
substantially larger than $e$. This is why we adopted the
iterative method described above.

However, using the same type of eccentricity expansion, one can
check the statement made earlier about the impact of the
$\theta=\phi$ assumption; the difference lies in the fact that in
this case the expansion at order $k$ is really of magnitude $e^k$.

\subsection{Recovery of the primary kinematic paramaters}\label{primaryparam}

The approximate WKBJ ordering which applies over nearly all the
wave propagation region suggests that Eqs.~(\ref{qcosgamma}) and
(\ref{qsingamma}) can be used in an iterative way to determine $e$
and $\gamma$. In practice, defining the $n$th order approximation
of $e$ and $\gamma$ as $e^{(n)}$ and $\gamma^{(n)}$, we use:

\begin{equation}\label{en}
ae^{(n)}=\frac{q\sin\gamma^{(n)}}{[df/da - d\gamma^{(n)}/da]},
\end{equation}

\noindent and

\begin{equation}\label{gamman}
\cos\gamma^{(n+1)}=\frac{1}{q}\frac{dae^{(n)}}{da}.
\end{equation}

\noindent Note that these equations are used alternatively and not
simultaneously: $\gamma^{(n)}$ gives $e^{(n)}$ from
Eq.~(\ref{en}), and $e^{(n)}$ in turn gives $\gamma^{(n+1)}$ from
Eq.~(\ref{gamman}). The iteration is started with
$\gamma^{(0)}=\pi/2$. Obviously, this fails inside the first
wavelength, as convergence is ensured only when $\gamma$ is not
substantially different from $\pi/2$.

These expressions involve derivatives of $f$ and $\gamma$ with
respect to $a$. We compute these derivatives by using a degree 5
polynomial interpolation of $f$ and a simple procedure for the
derivatives of $ae$ and $\gamma$. Furthermore, we must smooth $ae$
and $\gamma$ after each iteration in order to avoid propagation of
numerical noise through successive numerical derivatives. We stop
the iteration process when the results become stable.

\vskip 0.5in
\noindent {\bf ACKNOWLDEGMENTS:}

The research described in this publication was carried out at the
Jet Propulsion Laboratory, California Institute of Technology,
under a contract with the National Aeronautics and Space
Administration.  We are very grateful to Drs. M. Tiscareno and J.
Colwell for their careful reviews of the manuscript and excellent
suggestions.

\bibliographystyle{plainnat}

\bibliography{planeto-biblio8}

\begin{thebibliography}{37}
\providecommand{\natexlab}[1]{#1}
\providecommand{\url}[1]{\texttt{#1}}
\expandafter\ifx\csname urlstyle\endcsname\relax
  \providecommand{\doi}[1]{doi: #1}\else
  \providecommand{\doi}{doi: \begingroup \urlstyle{rm}\Url}\fi

\bibitem[{Borderies} and {Longaretti}(1987)]{BL87}
N.~{Borderies} and P.~Y. {Longaretti}.
\newblock {Description and behavior of streamlines in planetary rings}.
\newblock \emph{Icarus}, 72:\penalty0 593--603, December 1987.

\bibitem[{Borderies} et~al.(1983){Borderies}, {Goldreich}, and
  {Tremaine}]{BGT83}
N.~{Borderies}, P.~{Goldreich}, and S.~{Tremaine}.
\newblock {Perturbed particle disks}.
\newblock \emph{Icarus}, 55:\penalty0 124--132, July 1983.

\bibitem[{Borderies} et~al.(1985){Borderies}, {Goldreich}, and
  {Tremaine}]{BGT85}
N.~{Borderies}, P.~{Goldreich}, and S.~{Tremaine}.
\newblock {A granular flow model for dense planetary rings}.
\newblock \emph{Icarus}, 63:\penalty0 406--420, September 1985.

\bibitem[{Borderies} et~al.(1986){Borderies}, {Goldreich}, and
  {Tremaine}]{BGT86}
N.~{Borderies}, P.~{Goldreich}, and S.~{Tremaine}.
\newblock {Nonlinear density waves in planetary rings}.
\newblock \emph{Icarus}, 68:\penalty0 522--533, December 1986.

\bibitem[{Borderies-Rappaport} and {Longaretti}(1994)]{Borderies94}
N.~{Borderies-Rappaport} and P.-Y. {Longaretti}.
\newblock {Test particle motion around an oblate planet}.
\newblock \emph{Icarus}, 107:\penalty0 129--141, January 1994.

\bibitem[{Colwell} and {Esposito}(2007)]{Colwell07}
J.~E. {Colwell} and L.~W. {Esposito}.
\newblock {Density and Bending Waves in Saturn's Rings from Cassini UVIS Star
  Occultations}.
\newblock In \emph{AAS/Division for Planetary Sciences Meeting Abstracts},
  volume~39 of \emph{AAS/Division for Planetary Sciences Meeting Abstracts},
  page 26.06, October 2007.

\bibitem[{Colwell} et~al.(2006){Colwell}, {Esposito}, and {Srem{\v
  c}evi{\'c}}]{Colwell06}
J.~E. {Colwell}, L.~W. {Esposito}, and M.~{Srem{\v c}evi{\'c}}.
\newblock {Self-gravity wakes in Saturn's A ring measured by stellar
  occultations from Cassini}.
\newblock \emph{\grl}, 33:\penalty0 7201, April 2006.

\bibitem[{Colwell} et~al.(2007){Colwell}, {Esposito}, {Srem{\v c}evi{\'c}},
  {Stewart}, and {McClintock}]{Colwell07a}
J.~E. {Colwell}, L.~W. {Esposito}, M.~{Srem{\v c}evi{\'c}}, G.~R. {Stewart},
  and W.~E. {McClintock}.
\newblock {Self-gravity wakes and radial structure of Saturn's B ring}.
\newblock \emph{Icarus}, 190:\penalty0 127--144, September 2007.

\bibitem[{French} et~al.(1993){French}, {Nicholson}, {Cooke}, {Elliot},
  {Matthews}, {Perkovic}, {Tollestrup}, {Harvey}, {Chanover}, {Clark},
  {Dunham}, {Forrest}, {Harrington}, {Pipher}, {Brahic}, {Grenier}, {Roques},
  and {Arndt}]{French93}
R.~G. {French}, P.~D. {Nicholson}, M.~L. {Cooke}, J.~L. {Elliot},
  K.~{Matthews}, O.~{Perkovic}, E.~{Tollestrup}, P.~{Harvey}, N.~J. {Chanover},
  M.~A. {Clark}, E.~W. {Dunham}, W.~{Forrest}, J.~{Harrington}, J.~{Pipher},
  A.~{Brahic}, I.~{Grenier}, F.~{Roques}, and M.~{Arndt}.
\newblock {Geometry of the Saturn system from the 3 July 1989 occultation of 28
  SGR and Voyager observations}.
\newblock \emph{Icarus}, 103:\penalty0 163--214, June 1993.

\bibitem[{Goldreich} and {Tremaine}(1982)]{Goldreich82}
P.~{Goldreich} and S.~{Tremaine}.
\newblock {The dynamics of planetary rings}.
\newblock \emph{\araa}, 20:\penalty0 249--283, 1982.

\bibitem[{Hedman} et~al.(2007){Hedman}, {Nicholson}, {Salo}, {Wallis},
  {Buratti}, {Baines}, {Brown}, and {Clark}]{Hedman07}
M.~M. {Hedman}, P.~D. {Nicholson}, H.~{Salo}, B.~D. {Wallis}, B.~J. {Buratti},
  K.~H. {Baines}, R.~H. {Brown}, and R.~N. {Clark}.
\newblock {Self-Gravity Wake Structures in Saturn's A Ring Revealed by Cassini
  VIMS}.
\newblock \emph{\aj}, 133:\penalty0 2624--2629, June 2007.

\bibitem[{Jacobson} et~al.(2006){Jacobson}, {Antreasian}, {Bordi}, {Criddle},
  {Ionasescu}, {Jones}, {Mackenzie}, {Meek}, {Parcher}, {Pelletier}, {Owen},
  {Roth}, {Roundhill}, and {Stauch}]{Jacobson06}
R.~A. {Jacobson}, P.~G. {Antreasian}, J.~J. {Bordi}, K.~E. {Criddle},
  R.~{Ionasescu}, J.~B. {Jones}, R.~A. {Mackenzie}, M.~C. {Meek}, D.~{Parcher},
  F.~J. {Pelletier}, W.~M. {Owen}, Jr., D.~C. {Roth}, I.~M. {Roundhill}, and
  J.~R. {Stauch}.
\newblock {The Gravity Field of the Saturnian System from Satellite
  Observations and Spacecraft Tracking Data}.
\newblock \emph{\aj}, 132:\penalty0 2520--2526, December 2006.

\bibitem[{Karjalainen} and {Salo}(2004)]{KS04}
R.~{Karjalainen} and H.~{Salo}.
\newblock {Gravitational accretion of particles in Saturn's rings}.
\newblock \emph{Icarus}, 172:\penalty0 328--348, 2004.

\bibitem[{Kliore} et~al.(2004){Kliore}, {Anderson}, {Armstrong}, {Asmar},
  {Hamilton}, {Rappaport}, {Wahlquist}, {Ambrosini}, {Flasar}, {French},
  {Iess}, {Marouf}, and {Nagy}]{Kliore04}
A.~J. {Kliore}, J.~D. {Anderson}, J.~W. {Armstrong}, S.~W. {Asmar}, C.~L.
  {Hamilton}, N.~J. {Rappaport}, H.~D. {Wahlquist}, R.~{Ambrosini}, F.~M.
  {Flasar}, R.~G. {French}, L.~{Iess}, E.~A. {Marouf}, and A.~F. {Nagy}.
\newblock {Cassini Radio Science}.
\newblock \emph{Space Science Reviews}, 115:\penalty0 1--70, December 2004.

\bibitem[{Lewis} and {Stewart}(2000)]{LS00}
M.~C. {Lewis} and G.~R. {Stewart}.
\newblock {Effects of Self-Gravity on Wakes at the Encke Gap}.
\newblock In \emph{Bulletin of the American Astronomical Society}, pages
  1089--+, October 2000.

\bibitem[{Lewis} and {Stewart}(2005)]{Lewis05}
M.~C. {Lewis} and G.~R. {Stewart}.
\newblock {Expectations for Cassini observations of ring material with nearby
  moons}.
\newblock \emph{Icarus}, 178:\penalty0 124--143, November 2005.

\bibitem[{Longaretti}(1992)]{L92}
P.~{Longaretti}.
\newblock {Planetary Ring Dynamics: from Boltzmann's Equation to Celestial
  Mechanics}.
\newblock In D.~{Benest} and C.~{Froeschle}, editors, \emph{Interrelations
  Between Physics and Dynamics for Minor Bodies in the Solar System}, pages
  453--+, 1992.

\bibitem[{Longaretti}(1989)]{L89}
P.-Y. {Longaretti}.
\newblock {Saturn's main ring particle size distribution - an analytic
  approach}.
\newblock \emph{Icarus}, 81:\penalty0 51--73, 1989.

\bibitem[{Longaretti} and {Borderies}(1986)]{LB86}
P.-Y. {Longaretti} and N.~{Borderies}.
\newblock {Nonlinear study of the Mimas 5:3 density wave}.
\newblock \emph{Icarus}, 67:\penalty0 211--223, August 1986.

\bibitem[{Longaretti} and {Borderies}(1991)]{LB91}
P.-Y. {Longaretti} and N.~{Borderies}.
\newblock {Streamline formalism and ring orbit determination}.
\newblock \emph{Icarus}, 94:\penalty0 165--170, November 1991.

\bibitem[{Marouf} et~al.(1986){Marouf}, {Tyler}, and {Rosen}]{Marouf86}
E.~A. {Marouf}, G.~L. {Tyler}, and P.~A. {Rosen}.
\newblock {Profiling Saturn's rings by radio occultation}.
\newblock \emph{Icarus}, 68:\penalty0 120--166, October 1986.

\bibitem[{Nicholson} et~al.(1990){Nicholson}, {Cooke}, and {Pelton}]{NCP90}
P.~D. {Nicholson}, M.~L. {Cooke}, and E.~{Pelton}.
\newblock {An absolute radius scale for Saturn's rings}.
\newblock \emph{\aj}, 100:\penalty0 1339--1362, October 1990.

\bibitem[{Nicholson} et~al.(2008){Nicholson}, {Hedman}, {Clark}, {Showalter},
  {Cruikshank}, {Cuzzi}, {Filacchione}, {Capaccioni}, {Cerroni}, {Hansen},
  {Sicardy}, {Drossart}, {Brown}, {Buratti}, {Baines}, and
  {Coradini}]{Nicholson08}
P.~D. {Nicholson}, M.~M. {Hedman}, R.~N. {Clark}, M.~R. {Showalter}, D.~P.
  {Cruikshank}, J.~N. {Cuzzi}, G.~{Filacchione}, F.~{Capaccioni}, P.~{Cerroni},
  G.~B. {Hansen}, B.~{Sicardy}, P.~{Drossart}, R.~H. {Brown}, B.~J. {Buratti},
  K.~H. {Baines}, and A.~{Coradini}.
\newblock {A close look at Saturn's rings with Cassini VIMS}.
\newblock \emph{Icarus}, 193:\penalty0 182--212, January 2008.

\bibitem[{Rosen} et~al.(1991{\natexlab{a}}){Rosen}, {Tyler}, and
  {Marouf}]{Rosen91a}
P.~A. {Rosen}, G.~L. {Tyler}, and E.~A. {Marouf}.
\newblock {Resonance structures in Saturn's rings probed by radio occultation.
  I - Methods and examples}.
\newblock \emph{Icarus}, 93:\penalty0 3--24, September 1991{\natexlab{a}}.

\bibitem[{Rosen} et~al.(1991{\natexlab{b}}){Rosen}, {Tyler}, {Marouf}, and
  {Lissauer}]{Rosen91b}
P.~A. {Rosen}, G.~L. {Tyler}, E.~A. {Marouf}, and J.~J. {Lissauer}.
\newblock {Resonance structures in Saturn's rings probed by radio occultation.
  II - Results and interpretation}.
\newblock \emph{Icarus}, 93:\penalty0 25--44, September 1991{\natexlab{b}}.

\bibitem[{Showalter} and {Nicholson}(1990)]{Showalter90}
M.~R. {Showalter} and P.~D. {Nicholson}.
\newblock {Saturn's rings through a microscope - Particle size constraints from
  the Voyager PPS scan}.
\newblock \emph{Icarus}, 87:\penalty0 285--306, October 1990.

\bibitem[{Shu}(1984)]{S84}
F.~H. {Shu}.
\newblock {Waves in planetary rings}.
\newblock In R.~{Greenberg} and A.~{Brahic}, editors, \emph{Planetary Rings},
  pages 513--561, 1984.

\bibitem[{Shu} et~al.(1985{\natexlab{a}}){Shu}, {Dones}, {Lissauer}, {Yuan},
  and {Cuzzi}]{SDLYC85}
F.~H. {Shu}, L.~{Dones}, J.~J. {Lissauer}, C.~{Yuan}, and J.~N. {Cuzzi}.
\newblock {Nonlinear spiral density waves - Viscous damping}.
\newblock \emph{\apj}, 299:\penalty0 542--573, December 1985{\natexlab{a}}.

\bibitem[{Shu} et~al.(1985{\natexlab{b}}){Shu}, {Dones}, {Lissauer}, {Yuan},
  and {Cuzzi}]{Shu85}
F.~H. {Shu}, L.~{Dones}, J.~J. {Lissauer}, C.~{Yuan}, and J.~N. {Cuzzi}.
\newblock {Nonlinear spiral density waves - Viscous damping}.
\newblock \emph{\apj}, 299:\penalty0 542--573, December 1985{\natexlab{b}}.

\bibitem[{Shu} et~al.(1985{\natexlab{c}}){Shu}, {Yuan}, and {Lissauer}]{SYL85}
F.~H. {Shu}, C.~{Yuan}, and J.~J. {Lissauer}.
\newblock {Nonlinear spiral density waves - an inviscid theory}.
\newblock \emph{\apj}, 291:\penalty0 356--376, April 1985{\natexlab{c}}.

\bibitem[{Spilker} et~al.(2004){Spilker}, {Pilorz}, {Lane}, {Nelson},
  {Pollard}, and {Russell}]{S04}
L.~J. {Spilker}, S.~{Pilorz}, A.~L. {Lane}, R.~M. {Nelson}, B.~{Pollard}, and
  C.~T. {Russell}.
\newblock {Saturn A ring surface mass densities from spiral density wave
  dispersion behavior}.
\newblock \emph{Icarus}, 171:\penalty0 372--390, October 2004.

\bibitem[{Sremcevic} et~al.(2008){Sremcevic}, {Stewart}, {Albers}, {Colwell},
  and {Esposito}]{Stewart08b}
M.~{Sremcevic}, G.~R. {Stewart}, N.~{Albers}, J.~E. {Colwell}, and L.~W.
  {Esposito}.
\newblock {Density Waves in Saturn's Rings: Non-linear Dispersion and Moon
  Libration Effects}.
\newblock In \emph{AAS/Division of Dynamical Astronomy Meeting}, volume~39,
  page \#18.05, May 2008.

\bibitem[{Stewart} and {Sremcevic}(2008)]{Stewart08a}
G.~R. {Stewart} and M.~{Sremcevic}.
\newblock {Temporally Modulated Density Waves in Saturn's Rings}.
\newblock In \emph{AAS/Division of Dynamical Astronomy Meeting}, volume~39,
  page \#18.06, May 2008.

\bibitem[{Tiscareno} et~al.(2006){Tiscareno}, {Nicholson}, {Burns}, {Hedman},
  and {Porco}]{Tiscareno06}
M.~S. {Tiscareno}, P.~D. {Nicholson}, J.~A. {Burns}, M.~M. {Hedman}, and C.~C.
  {Porco}.
\newblock {Unravelling Temporal Variability in Saturn's Spiral Density Waves:
  Results and Predictions}.
\newblock \emph{\apjl}, 651:\penalty0 L65--L68, November 2006.

\bibitem[{Tiscareno} et~al.(2007){Tiscareno}, {Burns}, {Nicholson}, {Hedman},
  and {Porco}]{Tiscareno07}
M.~S. {Tiscareno}, J.~A. {Burns}, P.~D. {Nicholson}, M.~M. {Hedman}, and C.~C.
  {Porco}.
\newblock {Cassini imaging of Saturn's rings II. A wavelet technique for
  analysis of density waves and other radial structure in the rings}.
\newblock \emph{Icarus}, 189:\penalty0 14--34, July 2007.

\bibitem[{Weidenschilling} et~al.(1984){Weidenschilling}, {Chapman}, {Davis},
  and {Greenberg}]{Weidenetal84}
S.~J. {Weidenschilling}, C.~R. {Chapman}, D.~R. {Davis}, and R.~{Greenberg}.
\newblock {Ring particles - Collisional interactions and physical nature}.
\newblock In R.~{Greenberg} and A.~{Brahic}, editors, \emph{Planetary Rings.
  Univ.\ Arizona}, pages 367--415, 1984.

\bibitem[{Zebker} et~al.(1985){Zebker}, {Marouf}, and {Tyler}]{Zebkeretal85}
H.~A. {Zebker}, E.~A. {Marouf}, and G.~L. {Tyler}.
\newblock {Saturn's rings - Particle size distributions for thin layer model}.
\newblock \emph{Icarus}, 64, December 1985.

\end{thebibliography}

\clearpage

\begin{table}
\begin{center}
\begin{tabular}{lcccccccc} \hline Profile \# & 1 & 2 & 3 & 4 & 5 & 6 & 7 & 8 \\
\hline $m \phi$ & -29.477 & -160.081 & 176.166 & 44.460 & 122.805 & 16.740 & 60.007 & -9.538 \\
\hline Applied shifts (km) & 0 & $-0.5$ & $-1.0$ & $0.5$ & $1.0$ & $-1.0$ & $-1.5$ & $1.5$ \\
 Computed shifts (km) & 0 & $-0.500$ & $-1.001$ & $0.502$ & $1.002$ & $-0.980$ & $-1.441$ & $1.501$ \\
Differences (m) & $0$ & $0$ & $1$ & $2$ & $2$ & $20$ & $59$ & $1$\\
\hline Applied shifts (km) & 0 & $-0.5$ & $-1.0$ & $0.5$ & $1.0$ &
$-1.0$ & $-1.5$ & $1.5$ \\ Computed shifts (km) & 0 & $-0.519$ &
$-1.038$ & $0.460$ & $0.978$ & $-1.009$ & $-1.524$ & $1.424$ \\
Differences (m) & 0 & $19$ & $38$ & $40$ & $22$ & $9$ & $24$ & $76
$ \\ \hline
\end{tabular}
\end{center}
\caption{First line: Values of $m \phi$ ($m=4$) in degrees for
each of the eight profiles. Next three lines: Computed shifts of
the simulated profiles without noise compared to the shifts
applied in generating the simulated data without noise; absolute
differences between the applied shifts and the computed shifts.
Last three lines: Computed shifts of the noisy simulated profiles
compared to the shifts applied in generating the simulated data;
absolute differences between the applied shifts and the computed
shifts.} \label{tab1}
\end{table}

\clearpage
\begin{table}
\begin{center}
\begin{tabular}{llcccc} \hline Profile & Rev. & Date & $a_{res}$ & $\overline{\tau}$ & Radial shift \\ \hline
1 & 7 & May 3, 2005 & 132301.717 & 0.84 & 0 \\
2 & 7 & May 3, 2005 & 132301.717 & 0.72 & 1.550 \\
3 & 8 & May 21, 2005 & 132302.037 & 0.78 & 0.564 \\
4 & 8 & May 21, 2005 & 132302.037 & 0.77 & 1.449 \\
5 & 10 & June 26, 2005 & 132300.601 & 0.72 & 0.361\\
6 & 10 & June 26, 2005 & 132300.601 & 0.77 & 0.146 \\
7 & 12 & August 2, 2005 & 132299.480 & 0.72 & 0.830 \\
8 & 12 & August 2, 2005 & 132299.480 & 0.75 & 0.177 \\
\hline \end{tabular}
\end{center}
\caption{Location $a_{res}$ of the Mimas 5:3 resonance for the
four dates of each profile. Each date has an ingress and an egress
so we have eight profiles.  Average optical depth
$\overline{\tau}$ for each profile in an unperturbed region of 20
km inward of the Mimas 5:3 resonance.  Radial shifts obtained by
the procedure.} \label{tab2}
\end{table}

\clearpage

\begin{figure}[htb]
\centering
\includegraphics[scale=1.0]{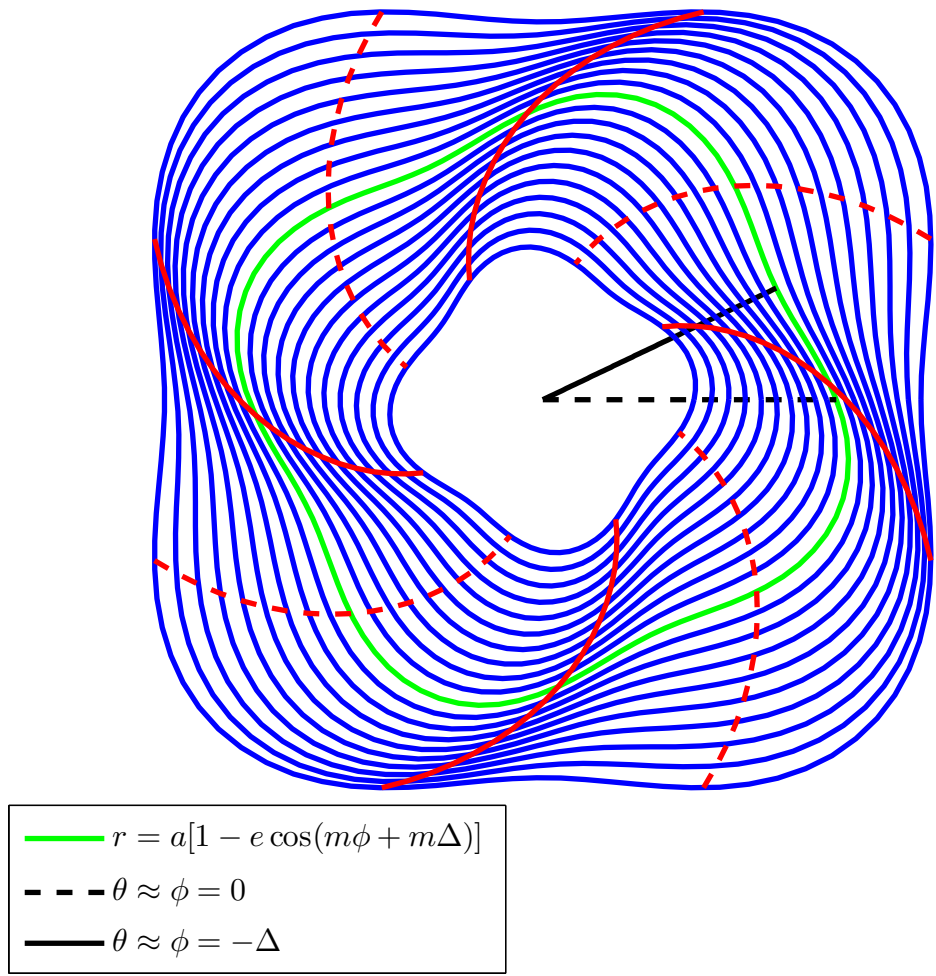}
\caption{\small Streamlines of a 4-lobe ($m=4$)density wave, such
as Mimas 5:3. The lag angle $\Delta$ provides the relative shift
in angle of one streamline to the next. The loci of constant
profile phase $\psi$ corresponding to the density maxima
($\psi=2\pi n$ with $n=$1 to 5, solid line) and minima ($\psi=\pi$
modulo $2\pi$, dashed line) are also shown to illustrate the
geometrical meaning of $\psi$. A streamline has been singled out,
to show the phase angle $\Delta$.}\label{fig0}
\end{figure}

\begin{figure}[htb]
\centering
\includegraphics[scale=1.0]{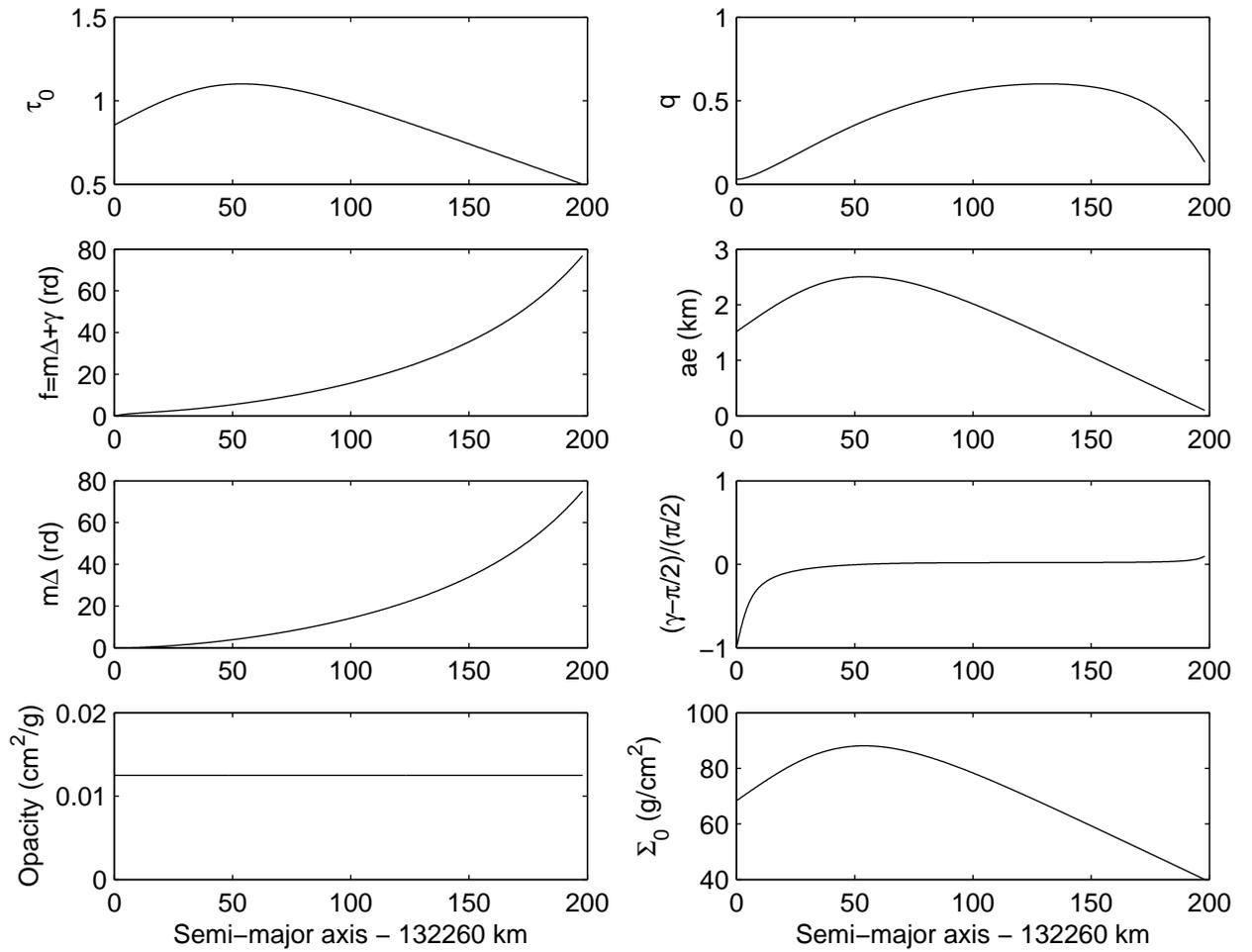}
\caption{\small Kinematic wave parameters for the simulated
profiles. These parameters were computed as described in section
\ref{simulation}. The assumed resonance radius is 132260 km.
}\label{fig1}
\end{figure}

\clearpage
\begin{figure}[htb]
\centering
\includegraphics[scale=1.0]{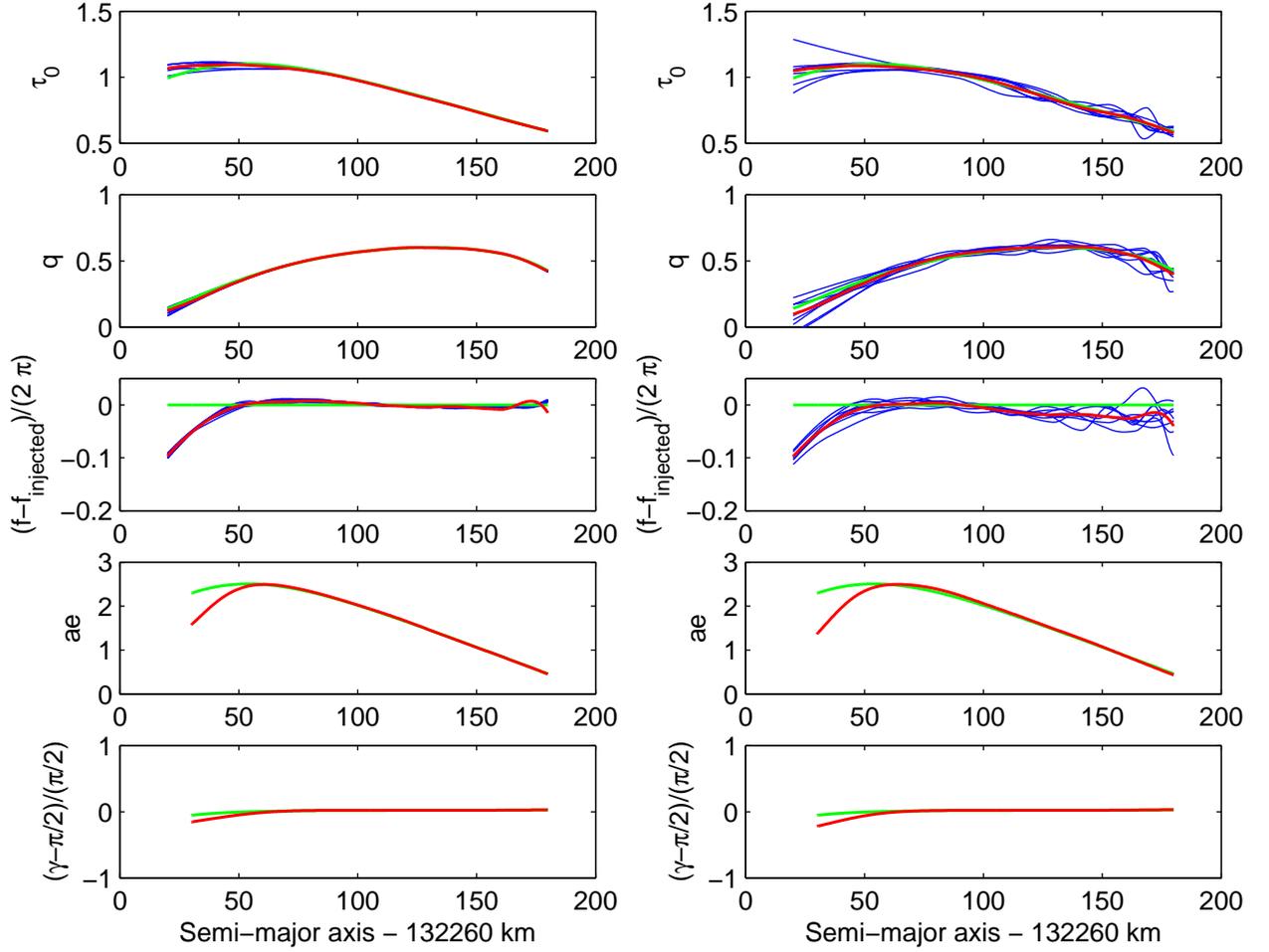}
\caption{\small Solution for the five wave kinematic parameters
after three passes for the simulation case without noise (left
panels) and with noise (right panels). The blue lines show the
solutions for the profiles taken independently from each other.
The red lines display the mean solution. The green lines refer to
the values injected in the simulation (see Fig.
\ref{fig1}).}\label{fig2}
\end{figure}

\clearpage
\begin{figure}[htb]
\centering
\includegraphics[scale=1.0]{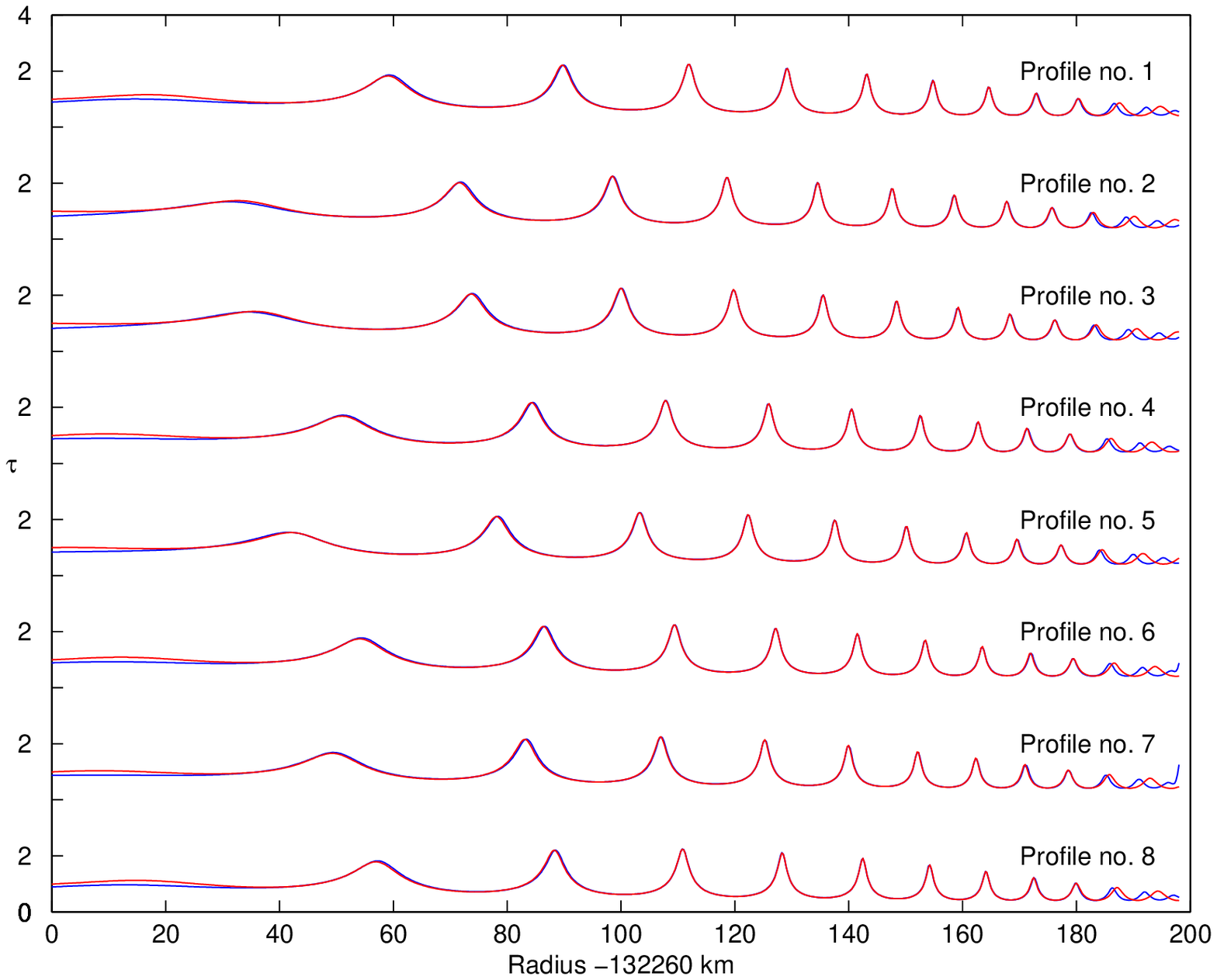}
\caption{\small Simulated profiles in blue for the case without
noise, and reconstructed proflles in red. For each profile, the
optical depth scale varies between 0 and 4. Intermediate tick
marks represent both the $\tau=0$ level of the next profile, and
the $\tau=4$ level of the previous one.}\label{fig3} \end{figure}

\clearpage
\begin{figure}[htb]
\centering
\includegraphics[scale=1.0]{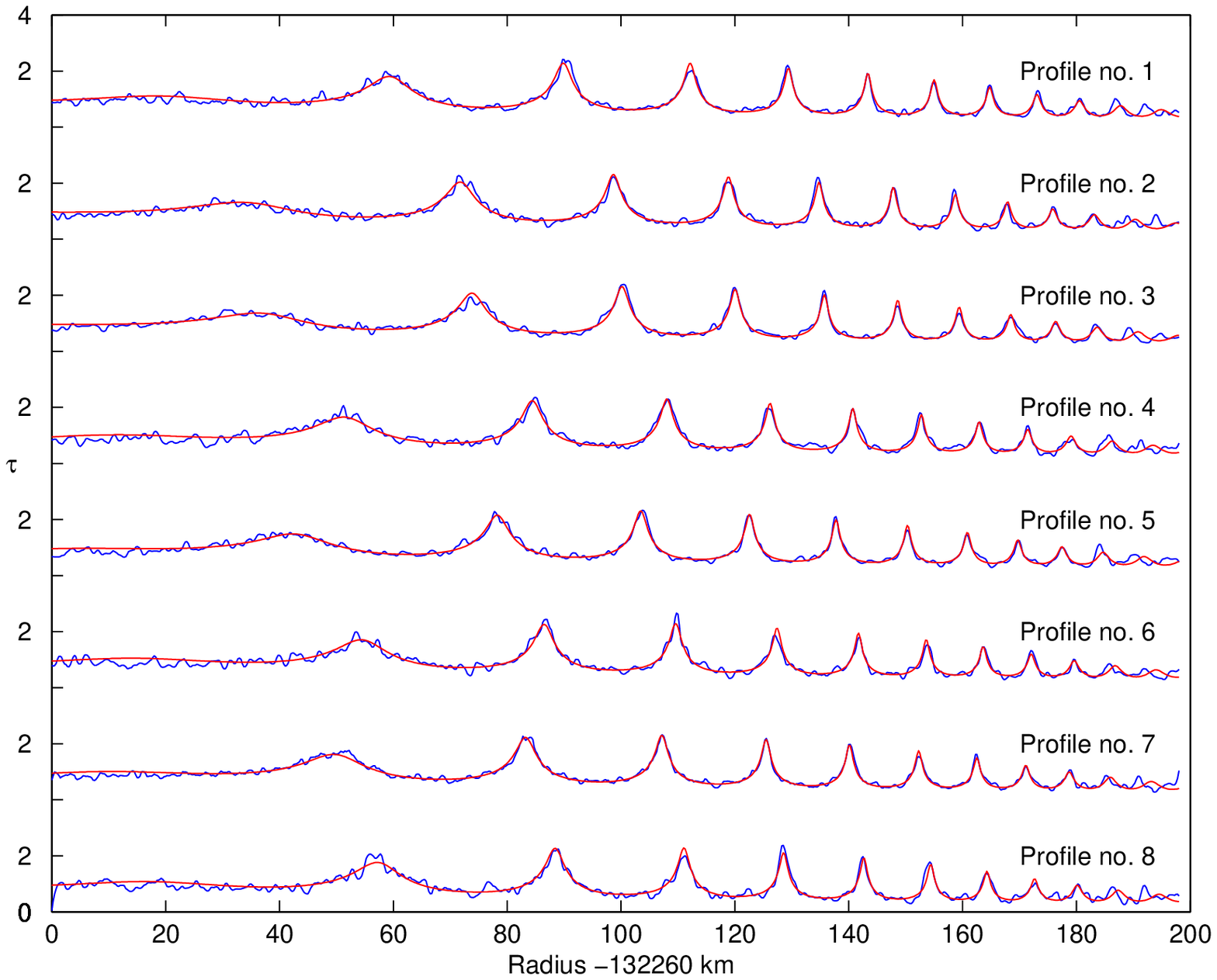}
\caption{\small Simulated profiles in blue for the case with
noise, and reconstructed proflles in red. For each profile, the
optical depth scale varies between 0 and 4. Intermediate tick
marks represent both the $\tau=0$ level of the next profile, and
the $\tau=4$ level of the previous one.}\label{fig4} \end{figure}

\clearpage
\begin{figure}[htb]
\centering
\includegraphics[scale=1.0]{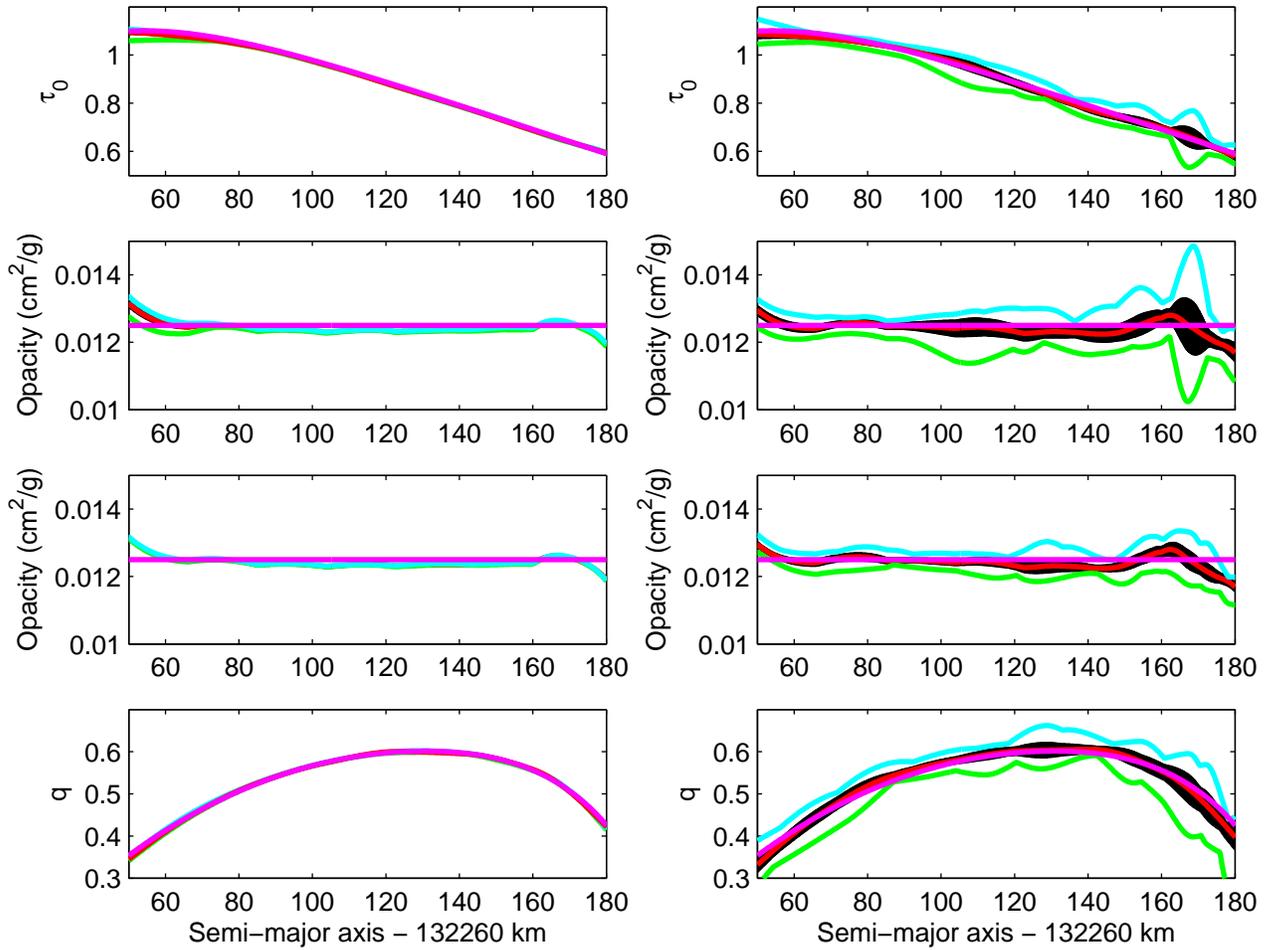}
\caption{\small Background optical depth, opacity, and
nonlinearity parameters for the simulated cases without noise
(left panels) and the cases with noise (right panels). In each
panel, the red line represents the solution, surrounded by its
error bars in black. The injected values used to generate the
simulated data are displayed by magenta lines. On the top and
bottom panels, the green and cyan lines show the minimum and
maximum values of the function obtained by analyzing the profiles
independently of each other up to the point where a mean solution
is determined. There are two panels for the opacity corresponding
to two ways of computing the error bars (see section
\ref{recopacity}). This section also explains how the green and
cyan lines were obtained.}\label{fig5}
\end{figure}

\clearpage
\begin{figure}[htb]
\centering
\includegraphics[scale=1.0]{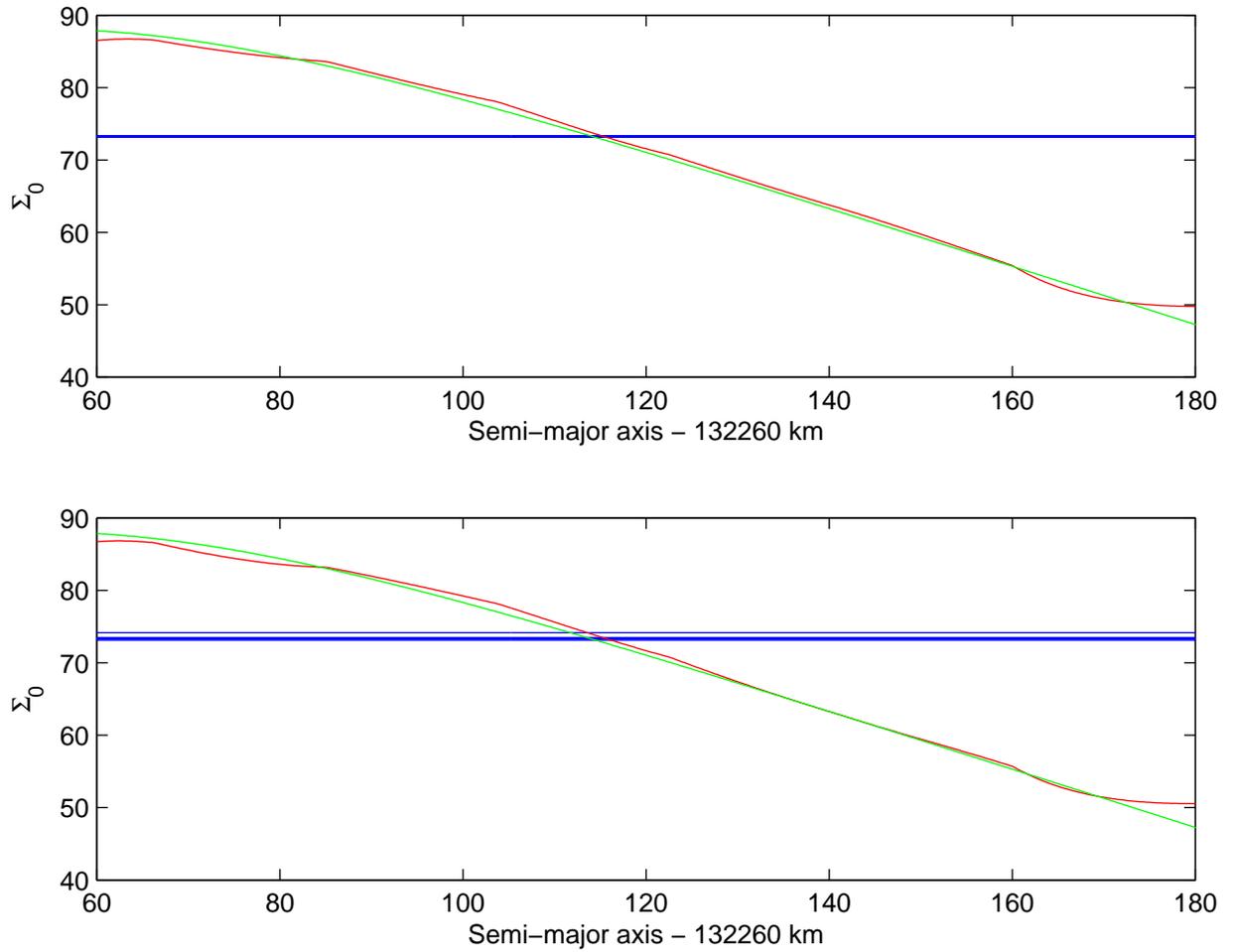}
\caption{\small For the simulation, surface density computed by a
fit to the linear theory (blue lines), computed by the nonlinear
theory (red), and injected in the simulation (green). The top
panel is for the case without noise, and the bottom panel is for
the case with noise.}\label{fig6}
\end{figure}

\clearpage
\begin{figure}[htb]
\centering
\includegraphics[scale=1.0]{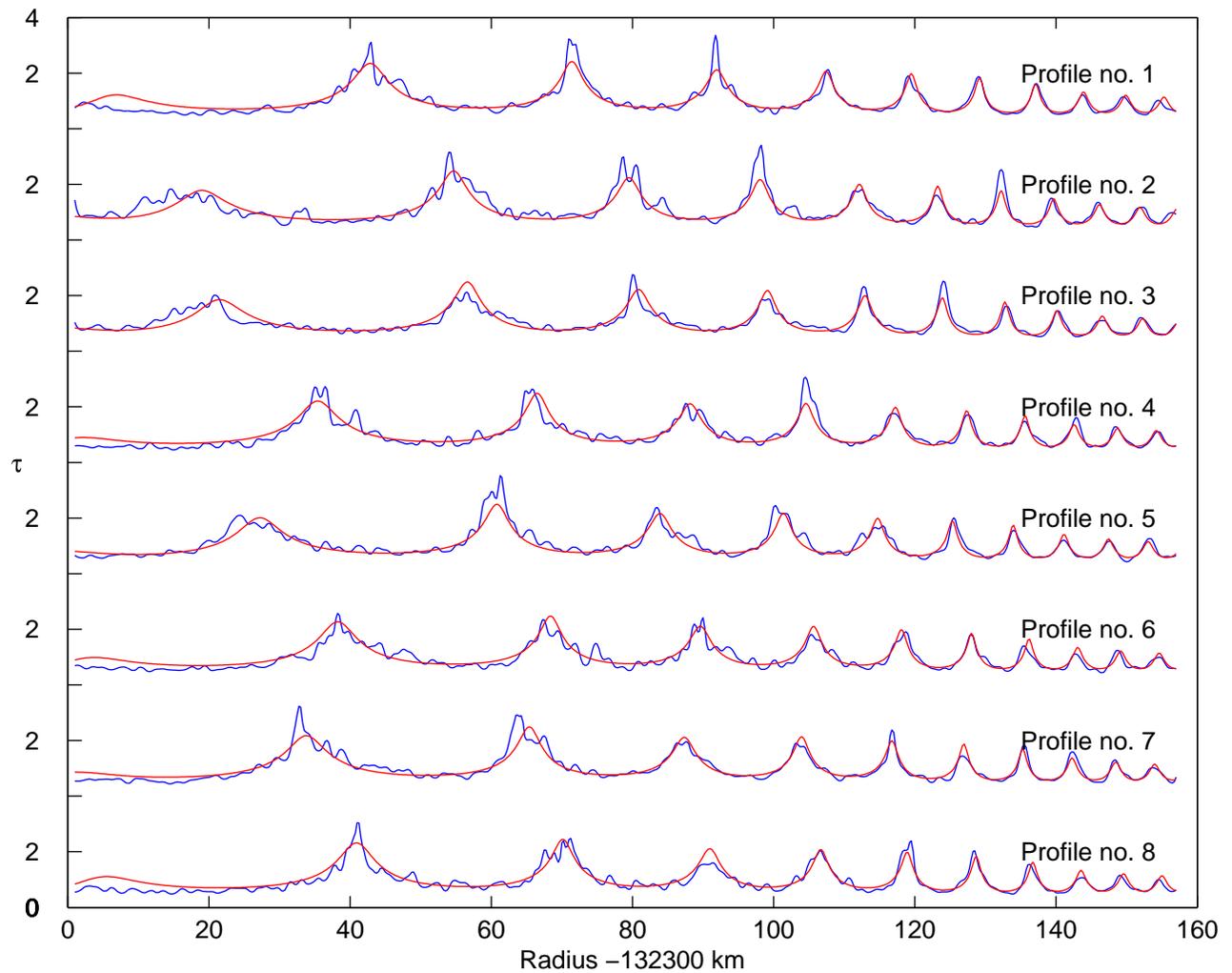}
\caption{\small Radio optical depth profiles for eight
occultations (in blue) of the Mimas 5:3 density wave and solution
after three passes (in red). For each profile, the vertical scale
goes from 0 to 4. Intermediate tick marks represent both the
$\tau=0$ level of the next profile, and the $\tau=4$ level of the
previous one. The resonance radius is 132301 km.}\label{fig7}
\end{figure}

\clearpage
\begin{figure}[htb]
\centering
\includegraphics[scale=1.0]{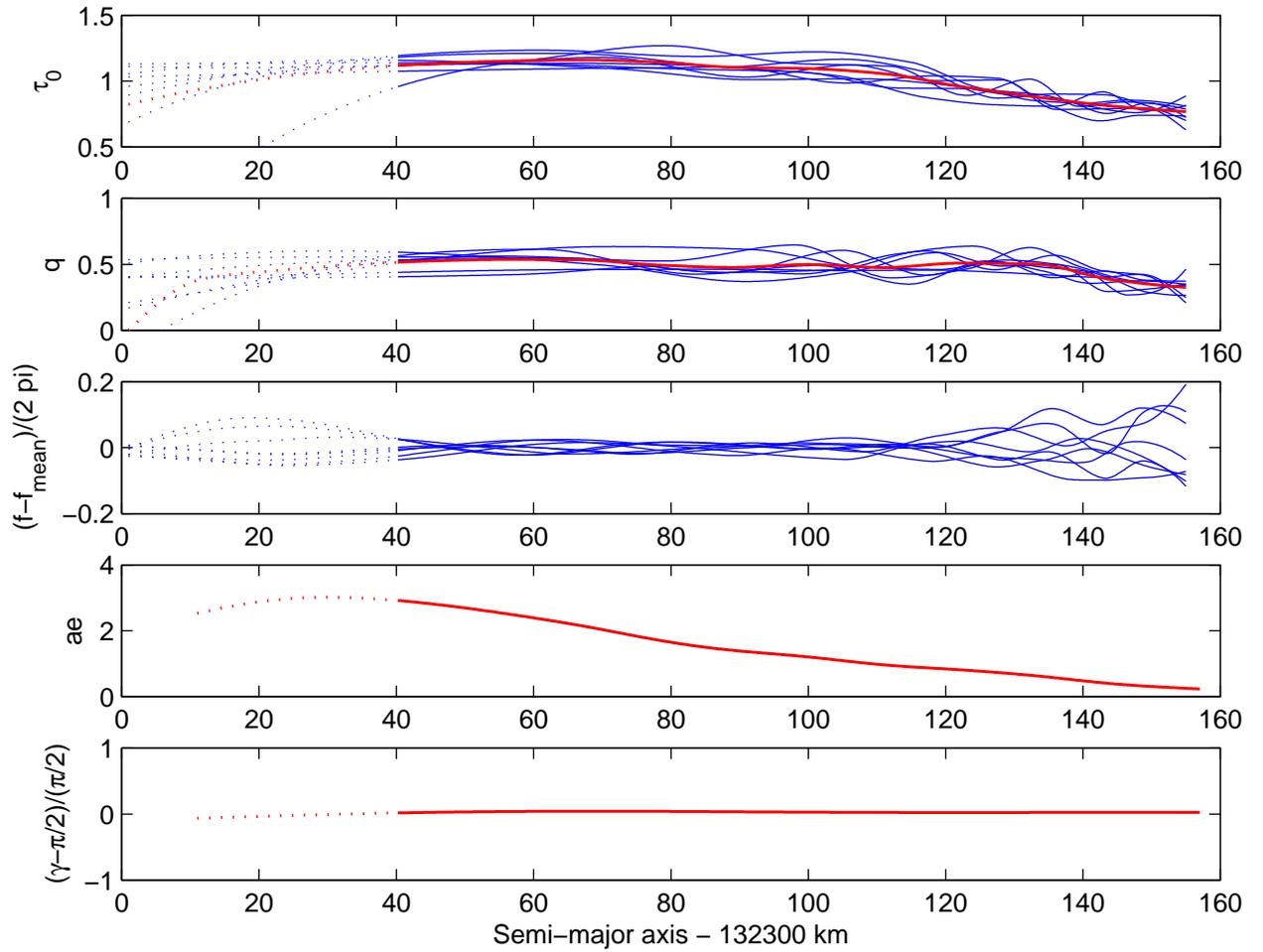}
\caption{\small Retrieved kinematic parameters of the Mimas 5:3
density wave. The kinematic parameters obtained by considering the
profiles independently from each other are shown in blue. The mean
solution is displayed in red. The dotted lines refer to the region
in which the parameters for at least one profile are
extrapolated.}\label{fig8}
\end{figure}

\clearpage
\begin{figure}[htb]
\centering
\includegraphics[scale=1.0]{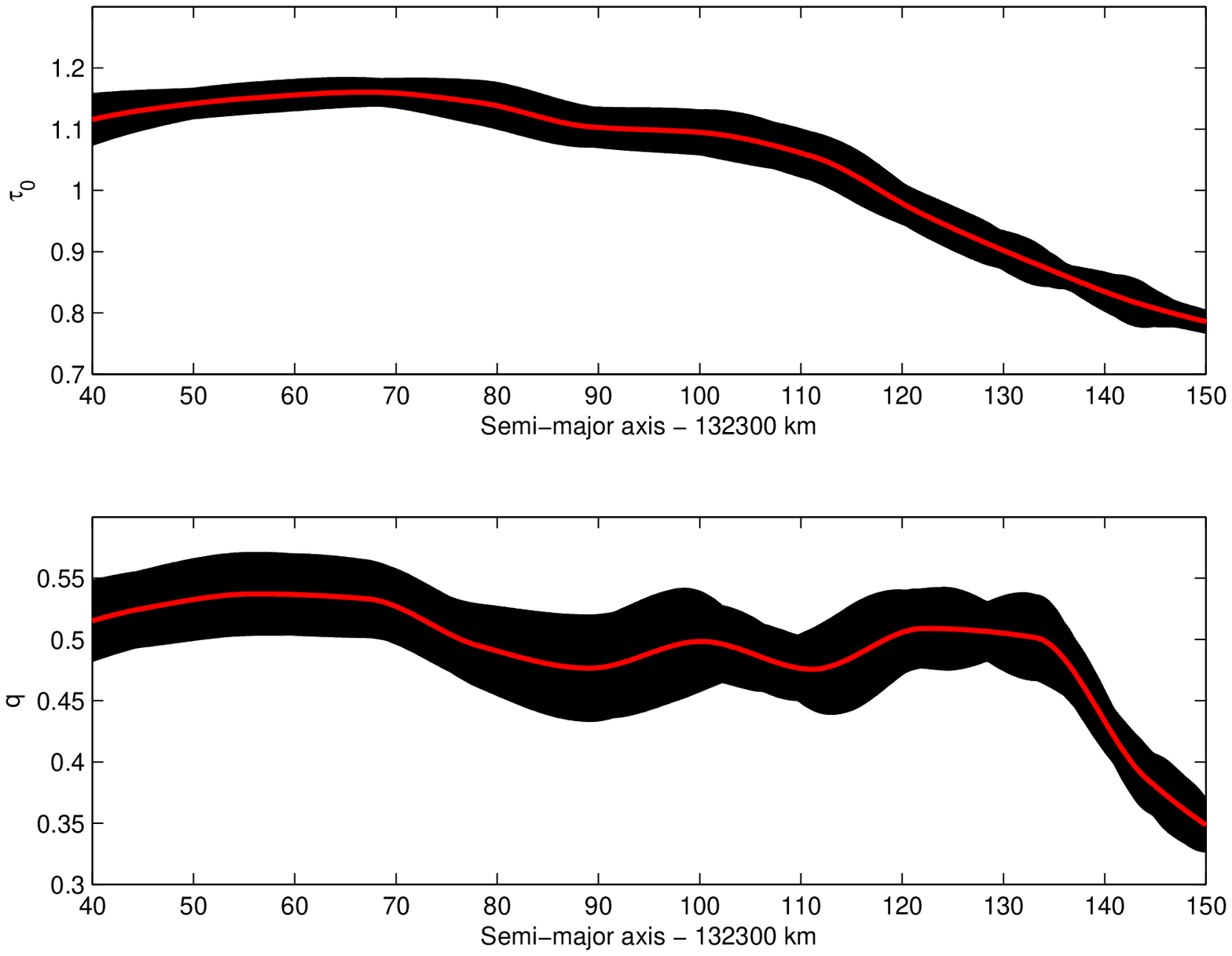}
\caption{\small Mean solution (in red) for the background optical
depth and the nonlinearity parameter surrounded by the error bars
in black, for the Mimas 5:3 density wave.}\label{fig9}
\end{figure}

\clearpage
\begin{figure}[htb]
\centering
\includegraphics[scale=1.0]{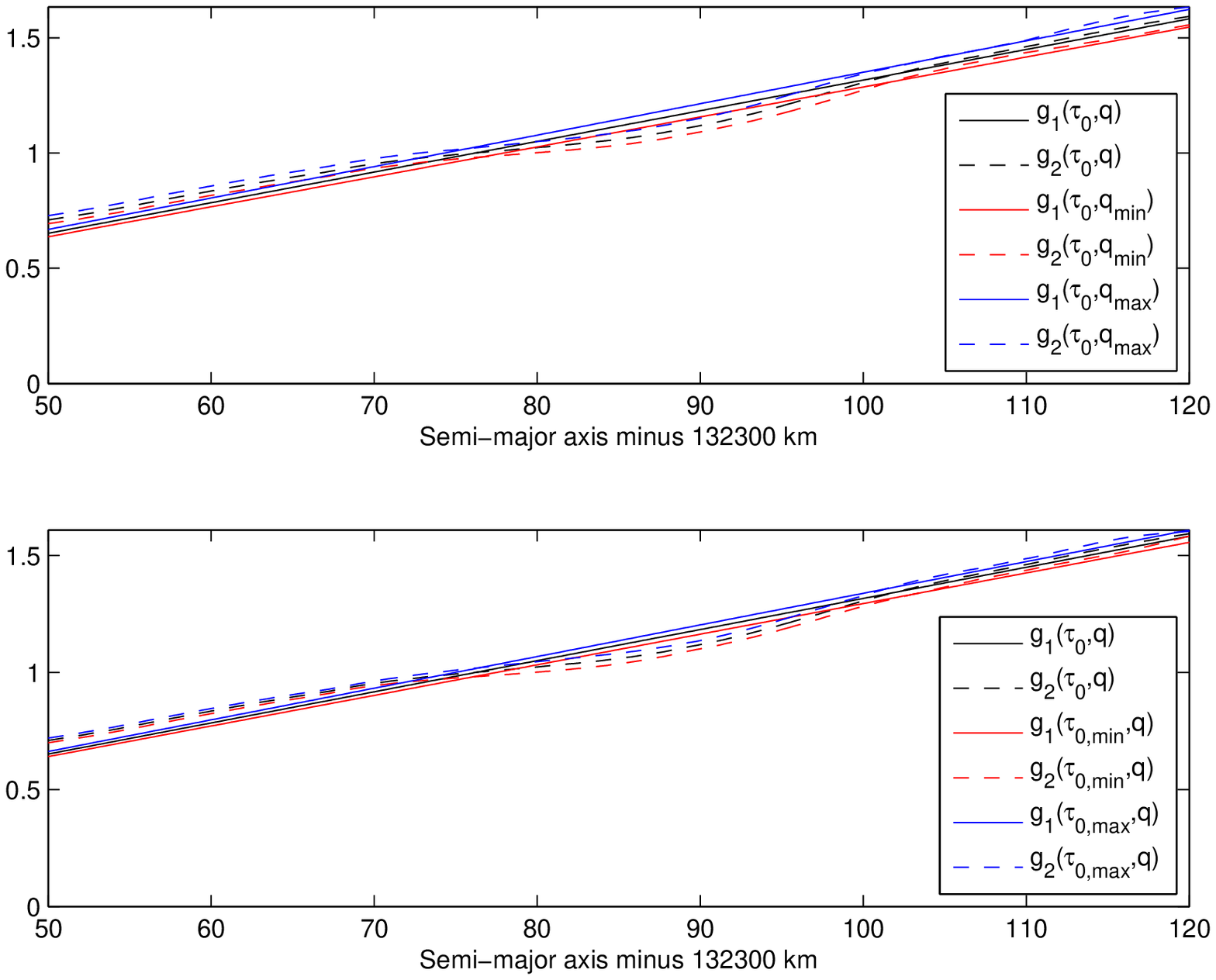}
\caption{\small Functions $g_1$ and $g_2$ given by Eqs.
(\ref{defg1}) and (\ref{defg2}) for various combination of
$\tau_0$ and $q$. Note that $\tau_{0,min}$, $\tau_{0,max}$,
$q_{min}$, and $q_{max}$ correspond to the values at the boundary
of the error bar areas, and not to the minimum or maximum values
of these variables obtained by treating the profiles
independently. }\label{fig10}
\end{figure}

\clearpage
\begin{figure}[htb]
\centering
\includegraphics[scale=1.0]{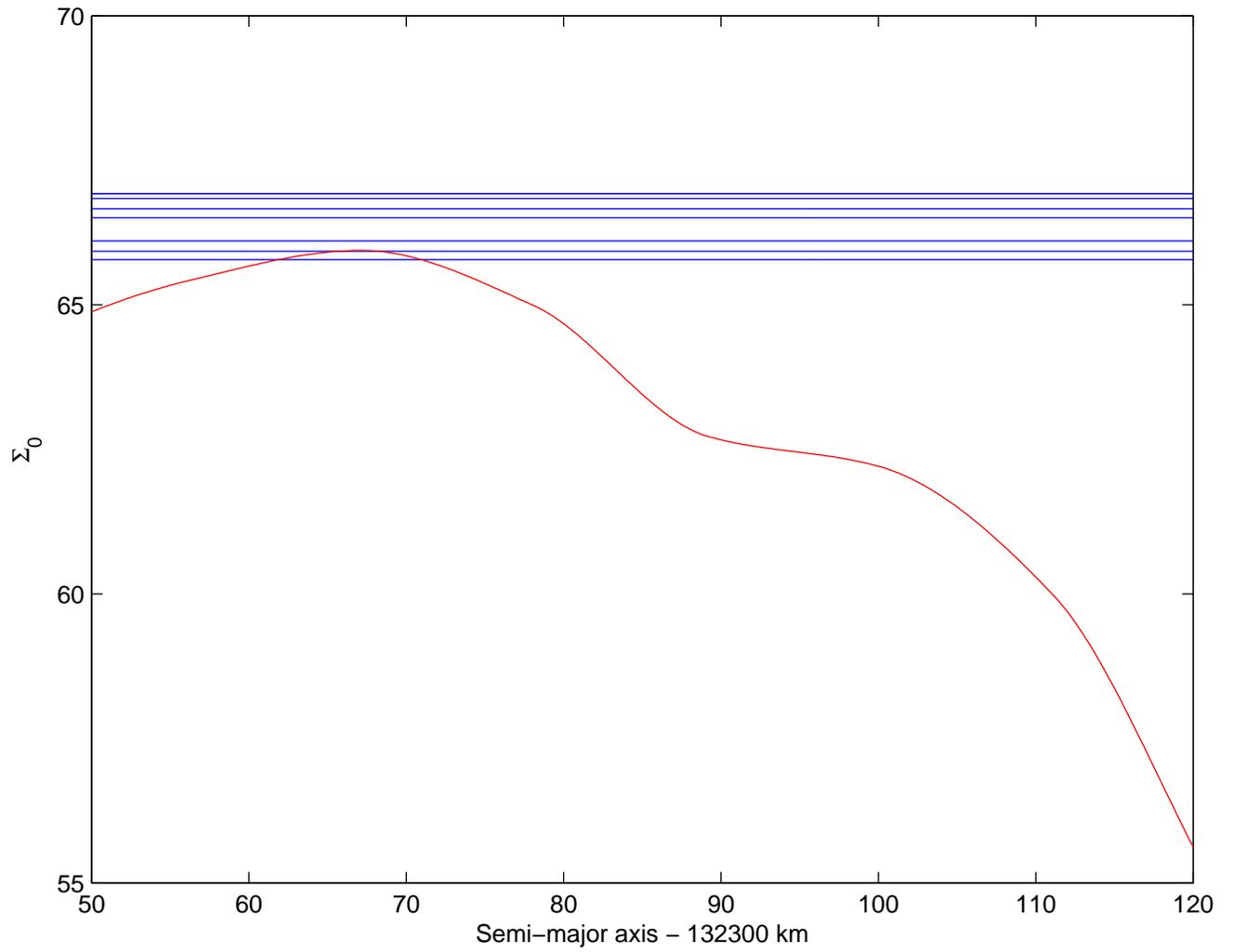}
\caption{\small Surface density from the nonlinear theory (in red)
and from the linear limit for each profile considered
independently (in blue) for the Mimas density wave.}\label{fig11}
\end{figure}

\clearpage
\begin{figure}[htb]
\centering
\includegraphics[scale=1.0]{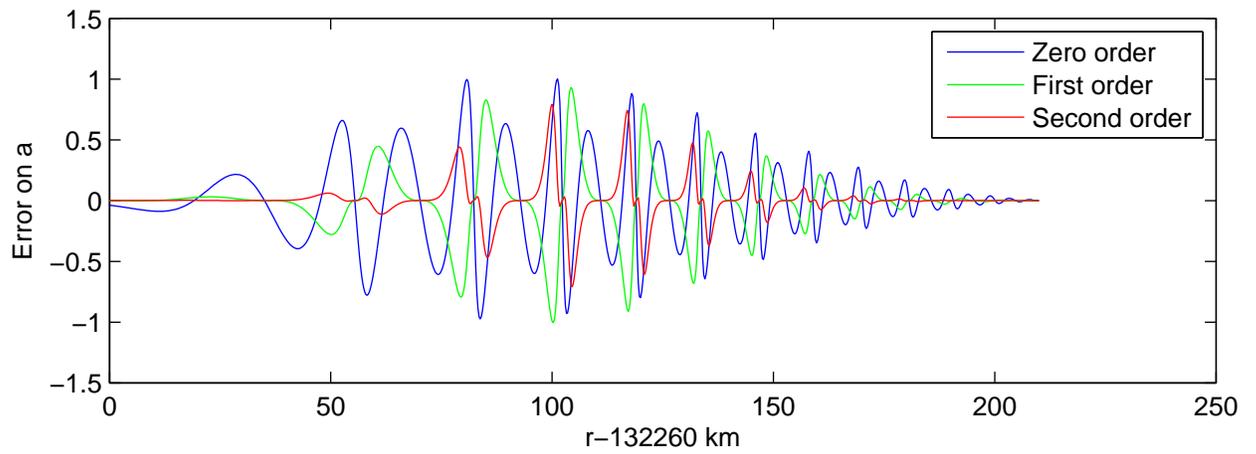}
\caption{\small Error between the approximate $a(r)$ obtained when
using the zeroth, first, and second order solutions and the actual
$a(r)$. }\label{fig12}
\end{figure}

\end{document}